\documentclass[twocolumn]{aastex631}

\usepackage{graphicx}	
\usepackage{amsmath}	
\usepackage{amssymb}	
\usepackage{bm}

\usepackage{hyperref}

\def\apjs{ApJS}

\def\apjl{ApJL}

\def\aap{A\&A}

\newcommand{\D}[2]{\frac{{\rm d} #2}{{\rm d} #1}}
\newcommand\bcdot{\,\bb{\cdot}\,}
\newcommand\btimes{\,\bb{\times}\,}

\newcommand\bb[1]{\mbox{\boldmath{$#1$}}}
\newcommand\grad{\bb{\nabla}}
\newcommand{\eb}{\hat{\bb{b}}}
\newcommand{\ez}{\hat{\bb{z}}}
\newcommand{\rmd}{{\rm d}}

\begin{document}

\title{Self-Similar Cosmic-Ray Transport in High-Resolution Magnetohydrodynamic Turbulence}
\shorttitle{CRs in turbulence}
\shortauthors{Kempski et al.}

\author[0009-0009-2144-3912]{Philipp Kempski}
\affiliation{Department of Astrophysical Sciences, Princeton University, Princeton, NJ 08544, USA}

\author[0000-0003-3806-8548]{Drummond B.~Fielding}
\affiliation{Department of Astronomy, Cornell University, Ithaca, NY 14853, USA}
\affiliation{Center for Cosmology and Particle Physics, Department of Physics, New York University, 726 Broadway, New York, NY 10003, USA}
\author{Eliot Quataert}
\affiliation{Department of Astrophysical Sciences, Princeton University, Princeton, NJ 08544, USA}

\author[0000-0002-0206-7271]{Robert J.~Ewart}
\affiliation{Department of Astrophysical Sciences, Princeton University, Princeton, NJ 08544, USA}

\author{Philipp Grete}
\affiliation{University of Hamburg, Hamburger Sternwarte, Gojenbergsweg 112, 21029 Hamburg, Germany}

\author[0000-0003-1676-6126]{Matthew W.~Kunz}
\affiliation{Department of Astrophysical Sciences, Princeton University, Princeton, NJ 08544, USA}
\affiliation{Princeton Plasma Physics Laboratory, PO Box 451, Princeton, NJ 08543, USA}

\author[0000-0001-7801-0362]{Alexander A. Philippov}
\affiliation{Department of Physics, University of Maryland, College Park, MD 20742, USA}

\author[0000-0001-5603-1832]{James Stone}
\affiliation{School of Natural Sciences, Institute for Advanced Study, 1 Einstein Drive, Princeton, NJ 08540, USA}

\correspondingauthor{Philipp Kempski}
\email{pkempski@princeton.edu}




\begin{abstract}
We study the propagation of cosmic rays (CRs) through a simulation of magnetohydrodynamic (MHD) turbulence at unprecedented resolution of $10{,}240^3$. We drive turbulence that is subsonic and super-Alfv\'enic, characterized by $\delta B_{\rm rms}/B_0=2$. The high resolution enables an extended inertial range such that the Alfv\'en scale $l_A$, where $\delta B (l_A)\approx B_0$, is well resolved. This allows us to properly capture how the cascade transitions from large amplitudes on large scales to small amplitudes on small scales. We find that sharp bends in the magnetic field are key mediators of particle transport even on small scales via resonant curvature scattering.  We further find that particle scattering in the turbulence shows strong hints of self-similarity: (1) the diffusion has weak energy dependence over almost two decades in particle energy and (2) the particles' random walk exhibits a broad power-law distribution of collision times such that the diffusion is dominated by the rarest, long-distance excursions. Our results suggest that large-amplitude MHD turbulence can provide efficient scattering over a wide range of CR energies and may help explain many CR observations above a $\sim$TeV: the flattening of the B/C spectrum, the hardening of CR primary spectra and the weak dependence of arrival anisotropy on CR energy.
\end{abstract}




\vspace{-55pt}

\section{Introduction}
Observations of cosmic rays (CRs) in the Milky Way show that these high-energy charged particles are efficiently confined by the Galactic magnetic field over a wide range of energies (\citealt{Strong2007}; \citealt{zweibel_micro}; \citealt{amato_blasi_18}; \citealt{Ruszkowski_Pfrommer2023}). From sub-GeV to PeV energies, CRs have residence times that are orders of magnitude longer than what would be predicted by ballistic escape from the Galaxy. The long lifetime is typically attributed to scattering off magnetic inhomogeneities  in the interstellar media (ISM) and halos of galaxies on scales comparable to the particle gyro-radii, with $\sim$GeV CRs scattering off AU-scale fluctuations and PeV CRs scattering off parsec-scale fluctuations. Thus, the existence of fluctuations on AU to parsec scales is critical for the  efficient confinement of GeV to PeV CRs.

A cascade of interstellar turbulence is a natural candidate for generating fluctuations on this broad range of scales. However, on small scales magnetized ISM turbulence is expected to form structures that are on average highly anisotropic (\citealt{gs95}), a feature that drastically suppresses CR scattering (\citealt{chandran_scattering}). A separate isotropic cascade of compressible MHD fast modes has been proposed as an alternative (\citealt{yan_lazarian_2004}; \citealt{yan_lazarian_2008}; \citealt{Xu2018}; \citealt{fornieri_2021}), but their strong damping due to steepening and non-ideal MHD effects is a significant challenge to this picture (\citealt{kq2022}). There is the additional possibility that CRs themselves generate the waves that scatter them through CR-driven instabilities (\citealt{kp69}; \citealt{skilling71}; \citealt{zweibel_micro};  \citealt{Holcomb2019}; 
\citealt{bai_mhd_pic}; \citealt{zweibel_2020}; \citealt{bai_2021}; \citealt{shalaby2021}; \citealt{Sun2024}; \citealt{Lemmerz2025}; \citealt{Schroer2025}), most notably the CR streaming instability.\footnote{It is also possible that instabilities driven by the thermal population of the plasma, or even charged astrophysical dust grains, provide some scattering (e.g., \citealt{Ji2022}; \citealt{Ewart2024}; \citealt{Reichherzer2025}).} However, these waves can likely only be excited by, and therefore scatter, CRs with energies less than a few hundred GeV (e.g., \citealt{kc71}; \citealt{farmer_goldreich}). As a result, self-generated waves cannot easily explain CR observations over the entire range of energies from GeV to PeV (\citealt{kq2022}; \citealt{hopkins_sc_et_problems}). 

Recent advancements in the field of MHD turbulence and particularly its intermittency (\citealt{alisa_dynamo}; \citealt{dong_2022}; \citealt{fielding_plasmoid}) have motivated a new class of CR propagation models that argue that CRs may be efficiently scattered in large-amplitude MHD turbulence by statistically rare regions in which the magnetic-field lines are bent sharply on scales of order the CR gyro-radius (\citealt{Lemoine2023}; \citealt{Kempski2023}), a mechanism we refer to as \textit{resonant-curvature scattering}. However, in their analyses both \cite{Lemoine2023} and \cite{Kempski2023} used turbulence that had either no net field or $\delta B_{\rm rms}/B_0 \gg 1$. As a result, it was not clear in this previous work whether curvature scattering also applies on small scales that are impacted by a dynamically relevant background field. In particular, in turbulence that satisfies $\delta B_{\rm rms}/B_0 < 1$
at the driving scale, high-curvature regions are significantly less abundant (\citealt{Yuen2020}).

In this Letter, we demonstrate that the resonant-curvature scattering of CRs proposed by \cite{Lemoine2023} and \cite{Kempski2023} remains efficient even on small scales. To this end, we performed an extremely high-resolution simulation of MHD turbulence ($10{,}240^3$; Fielding et al.,  in prep.)\footnote{Our simulation is the second-ever simulation of compressible MHD turbulence with resolution exceeding $10,000^{3}$ grid points. The first one, by \cite{Beattie2024}, studied the supersonic MHD dynamo at $10,080^3$.} in the mildly super-Alfv\'enic regime, with $\delta B_{\rm rms}/B_0\approx 2$, a regime of turbulence that may be common in the ISM of our Galaxy (\citealt{Ohno1993}; \citealt{Gaensler2005}; \citealt{jaffe2010}; \citealt{Haverkorn2015}). The high resolution of our simulation enables us to resolve how the turbulence transitions from large amplitudes on large scales to small amplitudes on small scales, and to study particle transport over an unprecedented range of energies.

The Letter is organized as follows. We describe the numerical methods in Section  \ref{sec:method} and present the derived transport coefficients from simulated particle trajectories in Section \ref{sec:diff_coeff}. We describe the physics governing the transport in Section \ref{sec:scattering} and discuss connections to CR observations in Section \ref{sec:discussion}. We summarize our results in Section \ref{sec:summary}.

\section{Numerical setup} \label{sec:method}
We use {\tt AthenaK} (\citealt{Stone2024}) to perform implicit large-eddy simulations (ILES) of isothermal ideal-MHD turbulence in a cubic box of size $L=1$ with resolutions ranging from $640^3$ to $10{,}240^3$. All simulations use a piecewise linear spatial reconstruction method, HLLD Riemann solver and constrained transport to ensure $\bb{\nabla \cdot B}=0$ \citep{StoneGardiner2009}. This is similar to the numerical method used by \cite{Grete:2023}, who showed that ILES simulations can closely match direct numerical simulations (DNS) when appropriate explicit diffusion coefficients are used. We generate turbulence using fully solenoidal driving in the wavenumber range $[2\pi/L, 6\pi/L]$ with random forcing that follows an Ornstein--Uhlenbeck process (\citealt{uhlenbeck_ornstein_1930}) with correlation time roughly equal to the largest eddy $(L/2)$ turnover time. 

\begin{figure}
    \includegraphics[width=\columnwidth]{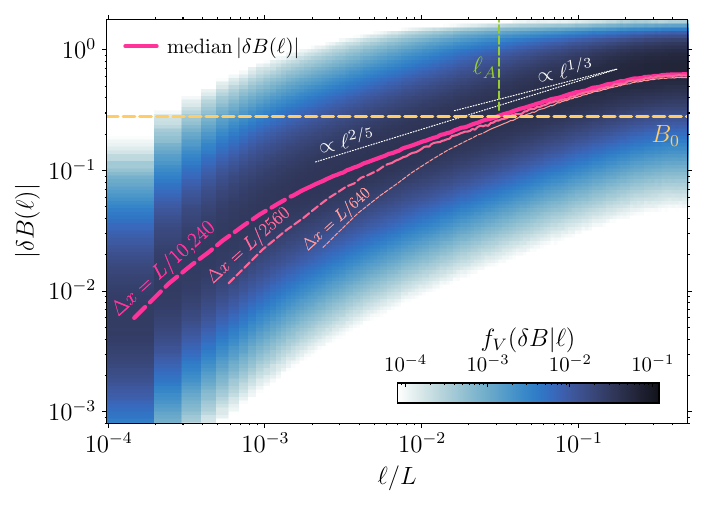}
    \caption{The shaded map shows the conditional probability density $f_V(\delta B|\ell)$---the fractional volume of pairs separated by a (isotropic) distance $\ell$ whose two-point increment has magnitude $|\delta B(\ell)|$---for our highest-resolution run ($L/\Delta x = 10{,}240$). The thick magenta curve traces the median $|\delta B|$ at each separation. Thinner magenta curves give the medians for lower-resolution simulations ($L/\Delta x = 640$ and $2560$). Their departure from the high-resolution median at small $\ell$ underscores the need for extreme resolution to capture both the super-Alfvénic ($|\delta B| \gtrsim B_0$) and sub-Alfvénic ($|\delta B| \lesssim B_0$) portions of the cascade within the resolved range. The median lines change to a dashed style at $\ell \leq 16\Delta x$ where numerical dissipation truncates the inertial range. The yellow horizontal dashed line marks the uniform guide field, $B_0$, while the green vertical dashed line denotes the Alfvén scale, $\ell_A \simeq L/30$ (${\approx}$300 grid cells), where the median $\delta B(\ell_A) = B_0$. White dotted segments indicate reference power-law scalings $\propto \ell^{2/5}$ and $\propto \ell^{1/3}$. Although the median fluctuation drops below $B_0$ for $\ell < \ell_A$, the PDF retains a substantial high-$|\delta B|$ tail: even an order of magnitude below $\ell_A$, a non-negligible fraction of pairs still satisfy $|\delta B| > B_0$. Cosmic rays with $r_L \ll \ell_A$ therefore encounter strongly perturbed fields despite the globally sub-Alfvénic character of the cascade.
    \label{fig:structure_function}
}
\end{figure}

The turbulence is driven subsonically with sonic Mach number $M_s=0.5$. The box is initially threaded by a uniform magnetic field $\bb{B_0} = B_0 \hat{\bb{z}}$ with initial plasma beta $\beta_0 \equiv 8\pi p_g/B_0^2=25$, where $p_g$ is the thermal gas pressure. The value of the initial magnetic field was chosen to ensure  that the turbulence saturates in the mildly super-Alfv\'enic regime with $\delta B_{\rm rms}/B_0 = 2$, where $\delta B_{\rm rms} = \langle (\bb{B}-\bb{B_0})^2\rangle^{1/2}$. In follow-up work we consider a variety of initial conditions at an intermediate resolution ($5120^3$) that result in a range of $\delta B_{\rm rms}/B_0$ in saturation. In this Letter, we focus on the properties of particle transport in turbulence characterized by $\delta B_{\rm rms}/B_0\approx 2$, which was run over a wide range of resolutions, including the unprecedentedly high resolution of $10{,}240^3$. This resolution allows the Alfv\'en scale, $l_A$, where $\delta B (l_A) \approx B_0$, to be resolved properly in the simulation (with $\gtrsim 300$ grid cells) and thus enables the first study of particle transport in MHD turbulence as it transitions from large amplitudes on large scales to small amplitudes on small scales. This is demonstrated in \autoref{fig:structure_function}, which shows the median $\delta B(\ell) \doteq |\bb{B}(x+\ell) - \bb{B}(x)|$ for simulations having resolutions of $10{,}240^3$, $2560^3$, and $640^3$, as well as the conditional probability density of $\delta B(\ell)$. Further properties of the turbulence are described in Fielding et al.~(in prep.).

The turbulence is evolved for approximately 6 outer-scale eddy-turnover times, which ensures statistical steady state in all quantities. After 6 turnover times, a large number of charged CRs are propagated through a static snapshot of the turbulent magnetic field using the Boris push method (\citealt{boris}) to integrate the equation of motion $ \rmd\bb{p}/\rmd t = q \bb{v}\btimes\bb{B}/c$, where $\bb{p}=\gamma m_0 \bb{v}$ is the CR momentum, $m_0$ is the rest mass of a proton, $\bb{v}$ is the CR velocity, and $q$ and $c$ are the charge and speed of light (both set to $1$).\footnote{As in \cite{Kempski2023}, we do not include the motional electric field, which for ISM conditions characterized by velocities $u \ll c$ is negligible compared to the magnetic field and does not significantly influence particle transport. This is also why the relativistic mass can be brought outside the time derivative in the particle equation of motion.} We evolve ultrarelativistic particles in 18 logarithmically spaced relativistic mass bins, with $r_L/L$ ranging from ${\approx}3 \times 10^{-4}$ to ${\approx}0.12$, with $10^{-4}$ particles per mass bin per cell.\footnote{In the $640^3$ and $2560^3$ simulations, we have an additional gyro-radius bin at $r_L/L\approx 0.17$.} This amounts to a total of ${\approx} 2 \times 10^9$ particles. For each mass bin, corresponding to a Lorentz factor $\gamma$ via $m=\gamma m_0$, we define the corresponding typical gyro-frequency $\Omega = q B_{\rm rms}/mc$ and gyro-radius $r_L = c/ \Omega$,  where $B_{\rm rms}$ is the root-mean-square magnetic-field strength. The particles are evolved for approximately $50$~box light-crossing times and we use the triangular-shaped-cloud method for interpolating the magnetic fields to the particle locations (\citealt{tsc_ref}). To study the extent to which field-line random walk affects the transport of CRs, we additionally integrate ${\sim}10^8$ particles that exactly follow field lines, i.e., particles that propagate ballistically along the local magnetic-field direction with a velocity $\bb{v} = c \eb$, where $\eb$ is the unit vector oriented along the local magnetic-field direction (thus, field-line particles do not undergo any pitch-angle scattering). From now on, we distinguish the particles that trace field lines from the remaining cosmic rays by referring to them as just ``field lines''. In this Letter, field-line diffusion therefore does not refer to the motion of magnetic-field lines by the turbulent velocity field, but instead refers to the spatial diffusion of the particles that move ballistically along field lines without any gyration.  

\begin{figure*}
    \includegraphics[width=\textwidth]{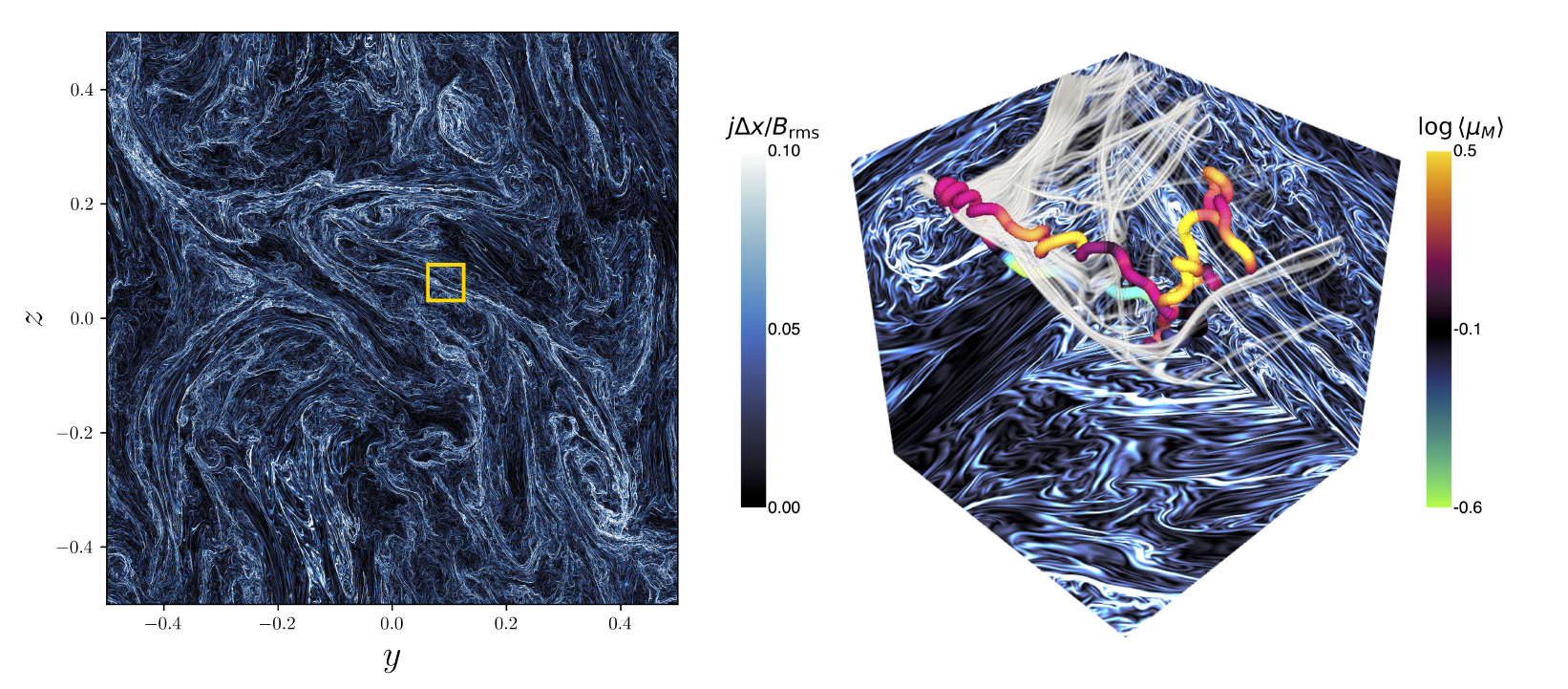}

  \caption{\textit{Left}: Two-dimensional slice of the magnitude of current density normalized by the grid spacing in the $10{,}240^3$ simulation characterized by $\delta B_{\rm rms}/B_0=2$ (Fielding et al., in prep.). \textit{Right:} Trajectory of a CR with $r_L/L \approx 7 \times 10^{-4}$ in a $640^3$ sub-domain (represented by the yellow square in the left panel) with side lengths approximately twice the Alfv\'en scale, $l_A$. The trajectory is color-coded using the logarithm of the particle's magnetic moment $\mu_{\rm M}$ averaged over a time window $2 \pi / \Omega$. The white lines show a collection of magnetic-field lines. The CR trajectory shows that there are various scattering mechanisms at play: $\mu_{\rm M}$-conserving magnetic mirroring where the field lines converge, as well as strong scattering events by large field-line curvature in which $\mu_{\rm M}$ changes rapidly (see also Figure~\ref{fig:mu_track}). The latter occur in regions where the magnetic field is weaker and more tangled.  \label{fig:j_particle}}
\end{figure*}

We show a slice of the current density from the $10{,}240^3$ simulation in the left panel of Figure \ref{fig:j_particle}, and the trajectory of a CR with $r_L/L \approx 7 \times 10^{-4}$, together with a collection of magnetic-field lines (white), in a $640^3$ sub-domain in the right panel. The trajectory is color-coded using the logarithm of the particle's magnetic moment $\mu_{\rm M} \equiv v_\perp^2 / 2B$ averaged over a time window $2 \pi / \Omega$. The CR undergoes both $\mu_{\rm M}$-conserving magnetic mirroring where the field lines converge, as well as $\mu_{\rm M}$-changing strong scattering events caused by regions of large field-line curvature. We discuss the nature of these scattering events in more detail in Section \ref{sec:scattering}.  

\section{Transport Coefficients} \label{sec:diff_coeff}
We now report the time evolution of our simulated CR and field-line transport coefficients, with a particular focus on their long-term behavior after reaching (approximate) steady state. We provide our physical interpretation of the results presented here in Section \ref{sec:scattering}. 

\begin{figure}
  \centering
    \includegraphics[width=0.49\textwidth]{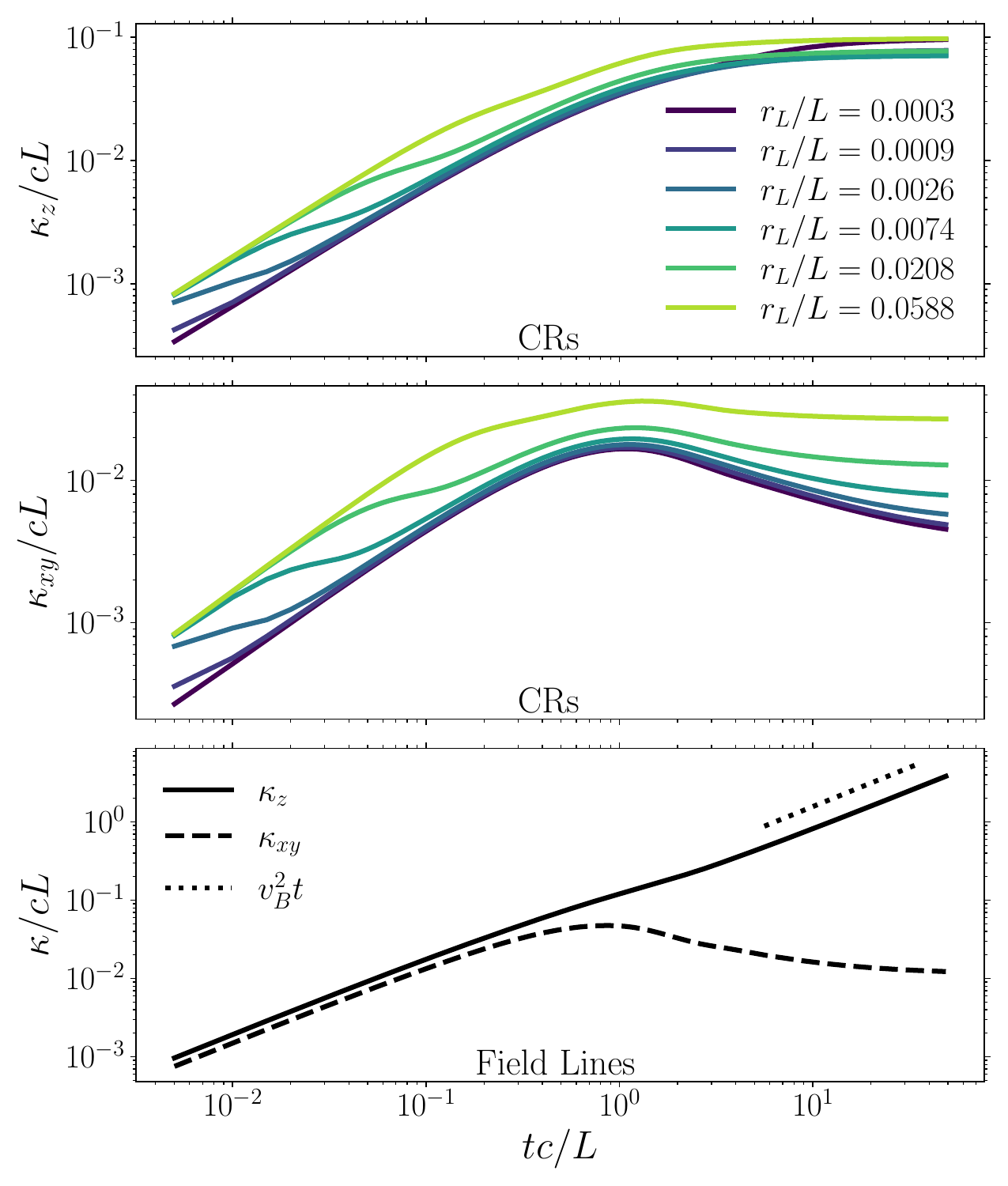}

  \caption{ Running diffusion coefficients in the $10{,}240^3$ simulation with $\delta B_{\rm rms}/B_0=2$. The diffusion coefficients are normalized by $cL$, where $c$ is the speed of light and $L$ is the size of the turbulent box. 
  \textit{Top}: running diffusion coefficient along the background guide field for particles with different gyro-radii $r_L$.
  \textit{Middle}: running diffusion coefficient of particles perpendicular to the guide field. 
  \textit{Bottom}: running diffusion coefficients perpendicular and parallel to the background guide field of magnetic field lines as traced by particles propagating ballistically along magnetic-field lines with velocity $\bb{v} = c \eb$ (this is described in detail at the end of Section~\ref{sec:method}). Because of the background guide field, the field lines show late-time ballistic behavior in the $z$-direction. \label{fig:running_diff}}
\end{figure}

Figure~\ref{fig:running_diff} shows the running spatial diffusion coefficients perpendicular and parallel to the background guide field for particles and field lines through the $10{,}240^3$ simulation domain. These coefficients are defined as $\kappa_z(t) \equiv \langle \Delta z ^2\rangle / 2 t$ and $\kappa_{xy}(t) \equiv \langle \Delta x^2 + \Delta y^2 \rangle / 4 t$, respectively, where $\langle \,\cdot\,\rangle$ denotes an average over all particles. We note that $t$ here represents the time as measured from the start of the particle/field-line integration and does not include the 6 largest-eddy turnover times over which the MHD fields were evolved. The top panel shows the evolution of $\kappa_z$ for a sample of mass bins (gyro-radii $r_L$), while the middle panel shows the evolution of $\kappa_{xy}$. After an initial ballistic phase on scales smaller than the scattering mean free path and the field-line coherence length, $\kappa_z$ reaches a steady-state diffusive plateau for all gyro-radii after ${\approx}10$ box light-crossing times. The evolution of $\kappa_{xy}$ shows a much more prolonged subdiffusive phase, especially at smaller gyro-radii. Subdiffusion perpendicular to the guide field is often thought of as being the product of compound diffusion: if particles diffuse along field lines, which themselves diffuse in the $x$-$y$ plane, this results in subdiffusion characterized by $\kappa_{xy}\sim t^{-1/2}$ (e.g., \citealt{Shalchi2019}). In our simulation, however, we instead find that field-line perpendicular transport is also characterized by an extended subdiffusive phase, as shown by the dashed line in the bottom panel of Figure~\ref{fig:running_diff}. Thus, the slower transport of CR particles in the $x$-$y$ plane is at least in part due to the fact that the large-scale geometry of magnetic-field lines in our mildly super-Alfv\'enic turbulence enforces subdiffusive propagation over long timescales. On shorter lengthscales, smaller than the turbulence driving scale, perpendicular transport of field lines is more efficient and comparable to transport along $z$ (solid line in the bottom panel), qualitatively consistent with the theory by \cite{NarayanMedvedev2001} for chaotic magnetic fields without a guide field. On short timescales, we find that $\sqrt{\langle \Delta x^2 + \Delta y^2 \rangle} \propto t$, which is different from the reported $t^{3/2}$ scaling for the perpendicular separation of field lines in sub-Alfv\'enic turbulence (\citealt{Lazarian2014}; \citealt{hu_2022}). However, here we do not track the separation between individual field lines, but instead their overall diffusion in the $x$-$y$ plane. In our super-Alfv\'enic turbulence there are many magnetic-field lines that have small pitch angles relative to the $x$-$y$ plane and it is not surprising that these field lines then produce the ${\propto}t$ scaling. On longer timescales, field-line transport along $z$ becomes much more efficient than transport in the $x$-$y$ plane due to the long-range correlations imposed by the background guide field. In particular, if we write the velocity of field lines as traced by particles with $\bb{v}=c\eb$ as $c \eb = c (\bb{B_0} + \bb{\delta B}(\bb{r}))/B(r) \sim c (\bb{B_0}/B_{\rm rms} + \bb{\delta B}(\bb{r})/B_{\rm rms})$, it becomes clear that field-line transport is a combination of ballistic propagation due to $\bb{B_0}$ and more chaotic (e.g., diffusive or subdiffusive) propagation due to $\bb{\delta B}$. On long timescales the ballistic component will dominate the transport even if $B_0 \ll B_{\rm rms}$, which is also what we see in the bottom panel. After an initial ballistic phase on scales smaller than the coherence length of the field, there is a brief sub-ballistic (superdiffusive) phase introduced by the fluctuating magnetic field. This phase does not last very long because the background guide field is relatively strong and for larger $t$ the $\kappa_z$ of field lines is consistent with ballistic propagation. To describe the late-time ballistic behavior of the field lines we define the effective ``magnetic-field velocity'',
\begin{equation} \label{eq:vb}
    v_B \equiv \sqrt{ \langle \Delta z^2 \rangle_{\rm FL}} / t,
\end{equation}
where the subscript ``${\rm FL}$" stands for field lines. In our simulations, we find that $v_B \approx 0.4c$, which is comparable to the ratio $B_0/B_{\rm rms}$ of our turbulence. This effective magnetic-field velocity is useful for understanding the relationship between $B$-field-aligned diffusion and $z$-aligned diffusion, especially at small particle energies. In particular, for fully magnetized particles moving along field lines like beads on wires, one expects that $\kappa_z$ is related to the diffusion coefficient calculated with respect to the local magnetic-field direction, 
\begin{equation}
    \kappa_\parallel = \Big\langle \Big( \int  \bb{v} \bcdot \eb\, \rmd t \Big)^2 \Big\rangle / (2t),
\end{equation}
as follows:
\begin{equation} \label{eq:kappaz_kappapar}
    \kappa_z = \kappa_\parallel (v_B/c)^2.
\end{equation}
Physically, a particle that is perfectly tied to a field line has to move a distance $\Delta \ell \sim \Delta z (c/v_B)$ to propagate a distance $\Delta z$ in the $z$-direction, with $\Delta \ell > \Delta z$ because the field lines are not aligned with $z$ but rather are tangled. Thus, $\kappa_z = \langle \Delta z^2 \rangle / 2t \approx (v_B/c)^2 \langle \Delta \ell^2 \rangle / 2t$, from whence follows \eqref{eq:kappaz_kappapar}.

\begin{figure}
  \centering
    \includegraphics[width=\columnwidth]{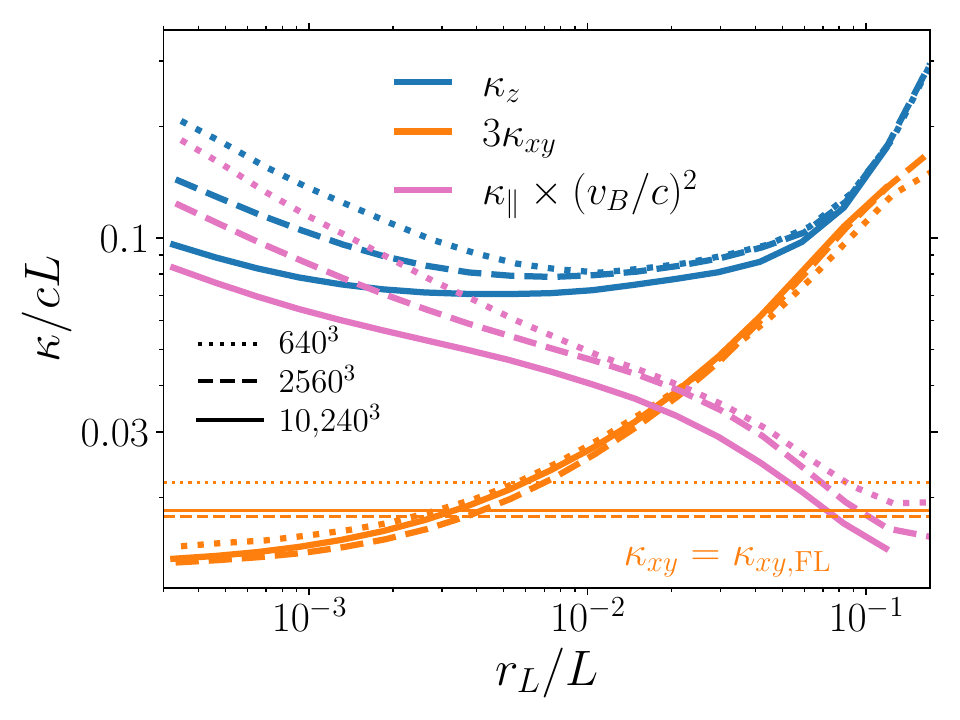}

  \caption{Diffusion coefficients parallel (blue) and perpendicular (orange) to the background guide field, and along the local magnetic field (pink) in our simulations with $\delta B_{\rm rms}/B_0=2$ at different resolutions. $\kappa_{xy}$ is multiplied by a factor of 3 for visualization purposes and we multiply $\kappa_\parallel$ by the effective late-time ballistic velocity of field lines (see bottom panel of Figure~\ref{fig:running_diff} and eq. \ref{eq:vb}) to compare with \eqref{eq:kappaz_kappapar}, which is the expected relationship for fully magnetized particles diffusing along field lines. Diffusion along $z$ is much more efficient than in the $x$-$y$ plane, demonstrating that the mean field renders transport extremely anisotropic despite the fact that $\delta B_{\rm rms} / B_0\approx 2$. The horizontal orange lines show the predicted perpendicular diffusion coefficients assuming particles are perfectly tied to field lines, using the late-time $\kappa_{xy}$ of magnetic-field lines (bottom panel of Figure~\ref{fig:running_diff}) and using that particle pitch angles satisfy on average $\langle |\mu| \rangle=0.5$. At small $r_L$, $xy$-diffusion thus appears to be well described by transport mediated by field-line random walk (FLRW), which turns out to be well converged with resolution. In contrast, $\kappa_z$, which dominates the overall diffusion rate, is not converged with resolution; at the highest resolution we find an extended range of gyro-radii where $\kappa_z$ has a weak dependence on particle energy. \label{fig:diff_coeffs}}
\end{figure}

Figure~\ref{fig:diff_coeffs} shows $\kappa_z$ (blue) and $\kappa_{xy}$ (orange, multiplied by $3$ for visualization purposes) evaluated at ${\approx}50$ box light-crossing times as functions of particle gyro-radius. Clearly, diffusion along the $z$ direction is significantly more efficient (by more than an order of magnitude at small rigidities) than in the $x$-$y$ plane, demonstrating that the background guide field renders transport very anisotropic despite the fact that $\delta B_{\rm rms} > B_0$. This is due to the very different behavior of field lines in the parallel and perpendicular directions (bottom panel of Figure~\ref{fig:running_diff}).  We also show the diffusion coefficient along the local magnetic-field direction, $\kappa_\parallel$ (pink), which is computed from the simulation by tracking the field-aligned displacement at each timestep, $\kappa_{\parallel} = \Big\langle \Big[ \sum_i ( \Delta \bb{r}_i \bcdot \eb_i) \Big]^2 \Big\rangle/(2t)$, where the angle bracket again denotes an average over particles, the index $i$ indicates the timestep, $\eb_i$ is the local magnetic-field direction at the particle location at timestep $i$, and $\Delta\bb{r}_i$ is the displacement at timestep $i$. We multiply $\kappa_\parallel$ by $(v_B/c)^2$ to compare with \eqref{eq:kappaz_kappapar}. Solid lines show results from the $10{,}240^3$ simulation, while dashed and dotted lines show the results at resolutions $2560^3$ and $640^3$, respectively. 

The orange curves in Figure \ref{fig:diff_coeffs} demonstrate that $\kappa_{xy}$ is remarkably converged with resolution, and reaches a plateau at small gyro-radii. The plateau is consistent with diffusion set by the rate at which field lines diffuse in the $x$-$y$ plane (horizontal orange lines), which also appears to be converged with resolution between the $2560^3$ and $10{,}240^3$ simulations. We discuss this behavior in more detail in Section~\ref{sec:scattering}. Diffusion in the $z$-direction (blue curves) shows more interesting behavior and a significant dependence on resolution: as resolution is increased, the energy (gyro-radius) of the best-confined (smallest $\kappa_z$) particle decreases. At the highest resolution we begin to see an extended range of particle gyro-radii for which $\kappa_z$ exhibits a weak dependence on energy, which suggests that we are seeing the first signs of an ``inertial range" in particle transport. Remarkably, in the $10{,}240^3$ simulation $\kappa_z$ is almost constant over almost 2 orders of magnitude in particle energy.   Figure~\ref{fig:diff_coeffs} further shows that $\kappa_\parallel (v_B/c)^2$ is comparable in value, and shows similar trends, to $\kappa_z$ (see \eqref{eq:kappaz_kappapar}) only for the smallest gyro-radii. For larger $r_L$, the pink and blue curves deviate from each other, suggesting that the particles become progressively less magnetized at higher energies, and their motion is a combination of magnetized and unmagnetized motion \citep{Kempski2023, Lemoine2023, Lubke2025}. We note that $\kappa_\parallel$ is also not converged with resolution, and shows appreciable flattening in the range $10^{-3}< r_L/L<10^{-2}$ as resolution is increased. As further explained in Section~\ref{sec:scattering}, the increased scattering along field lines at higher resolutions is key for explaining the resolution dependence of $\kappa_z$.  We speculate that this increased scattering along field lines is due to the fact that the width of the PDF of magnetic-field-line curvature increases with increasing resolution (Figure~\ref{fig:pdf_K_resolution}).

\section{Resonant-curvature scattering and L\'evy Flights} \label{sec:scattering}

\begin{figure}
  \centering
    \includegraphics[width=0.49\textwidth]{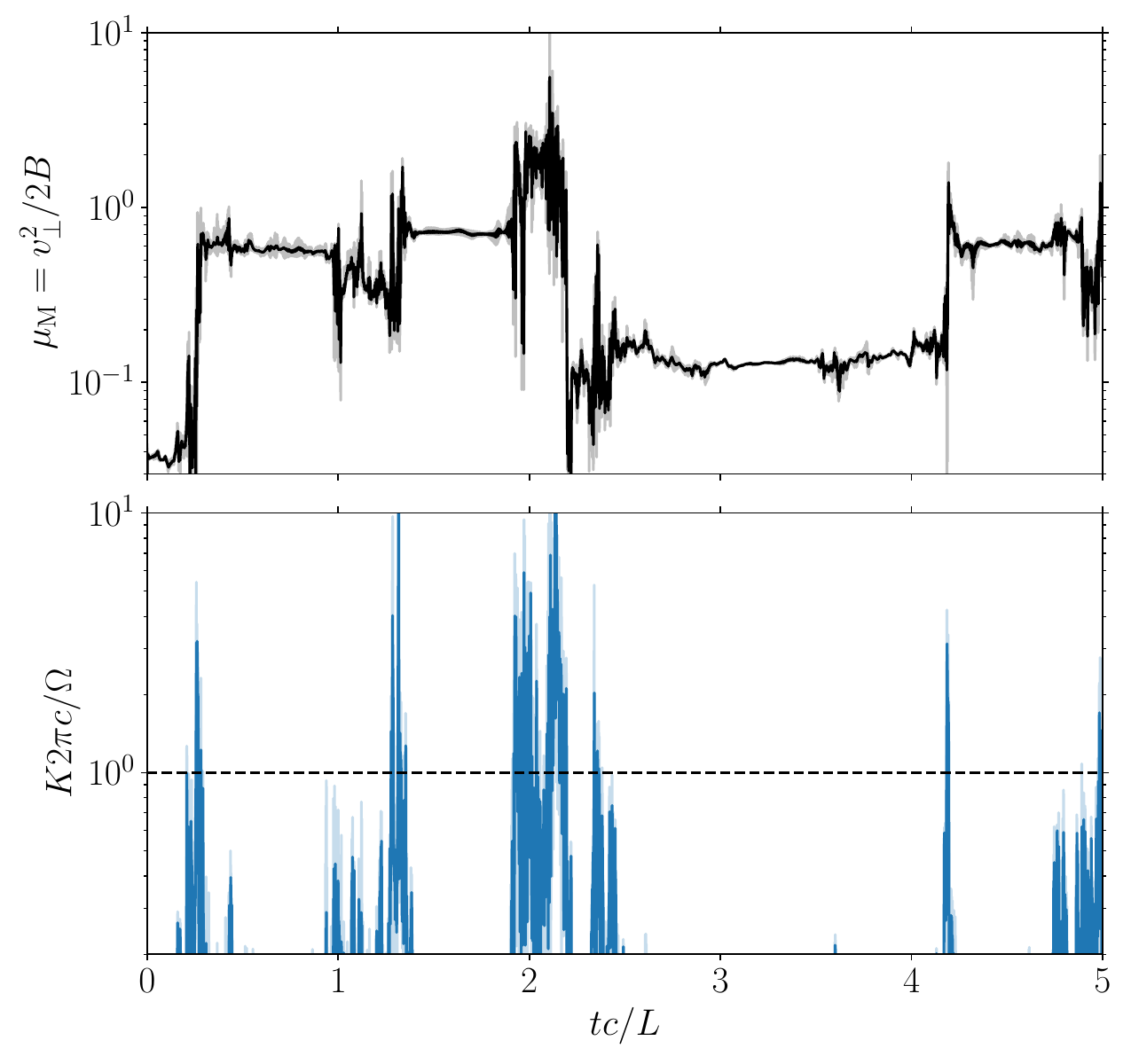}

  \caption{ Track of a particle with gyro-radius $r_L/L\approx 3 \times 10^{-4}$. Dark lines show running averages of tracked quantities over a window $2 \pi / \Omega$, faint lines show instantaneous values. Changes in the particle magnetic moment (top) are very abrupt and occur when the particle experiences regions of large field-line curvature, $K\equiv |\eb\bcdot\grad\eb|$, i.e., when $ K 2 \pi c/\Omega  \gtrsim 1$ (bottom). \label{fig:mu_track}}
\end{figure}

In Figure~\ref{fig:mu_track}, we show an example of data from a tracked particle that elucidates the primary scattering mechanism of small-$r_L$ particles. The track shows the history of a particle with $r_L \approx 3 \times 10^{-4} L$. From the top panel we see that the particle's magnetic moment, $\mu_{\rm M}$, changes in a step-like fashion, showing discrete large jumps (scattering events) interspersed with extended periods where it is roughly constant. The scattering events occur at times when the particle samples regions of resonant curvature, i.e., when the magnetic-field-line curvature $K \equiv \eb\bcdot \grad\eb$ locally satisfies $K 2 \pi c/\Omega \gtrsim 1$ (bottom panel), as previously demonstrated in \cite{Lemoine2023} and \cite{Kempski2023}. 

\begin{figure}
  \centering
            \begin{minipage}[b]{\textwidth}
\includegraphics[width=0.45\textwidth]{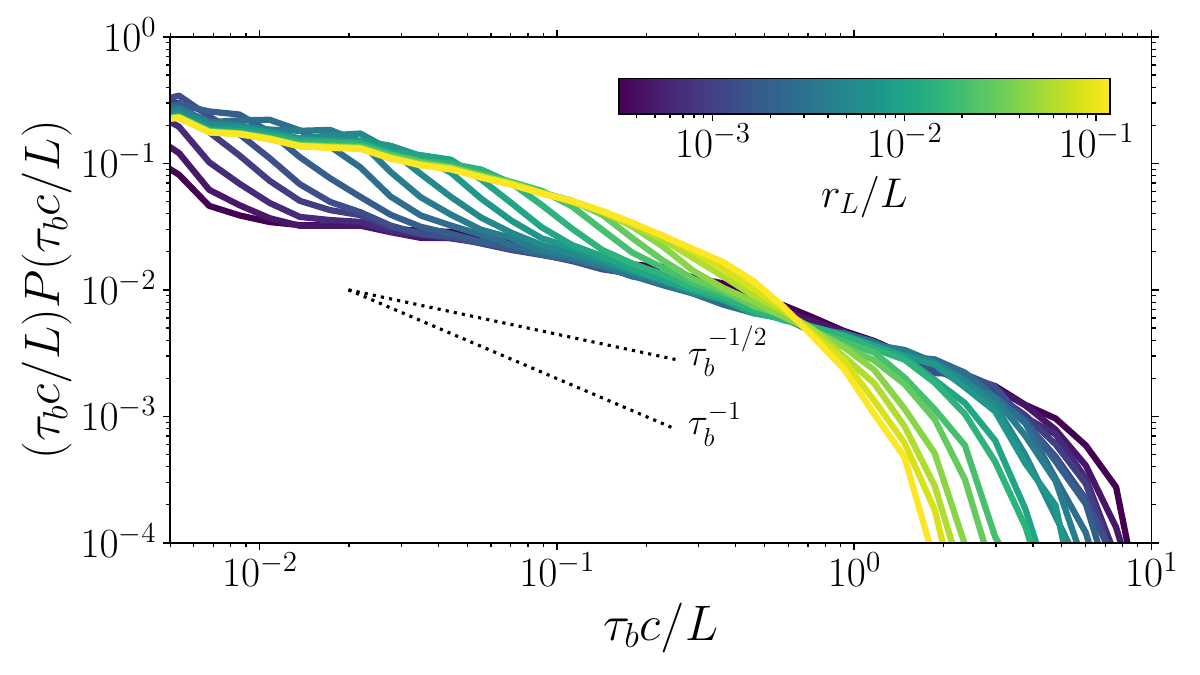}
        \end{minipage} 
        \vspace{-0.1in}
        \begin{minipage}[b]{\textwidth}
    \includegraphics[width=0.45\textwidth]{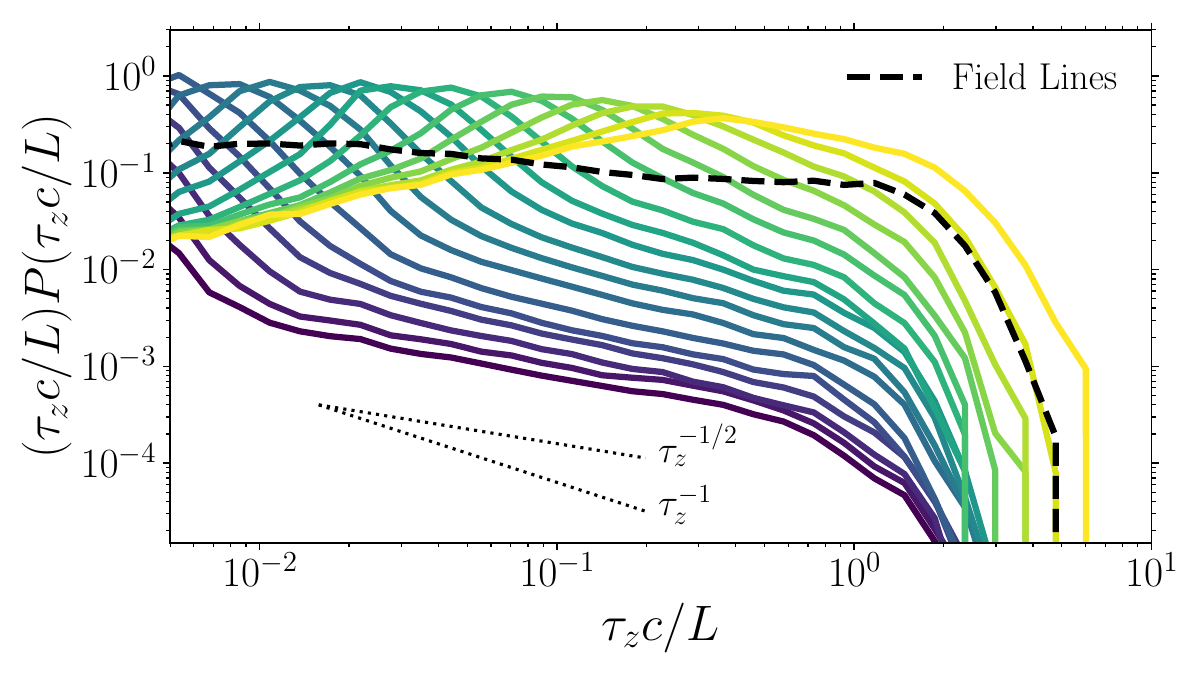}
        \end{minipage} 
        \vspace{-0.1in}

         \begin{minipage}[b]{\textwidth} \includegraphics[width=0.45\textwidth]{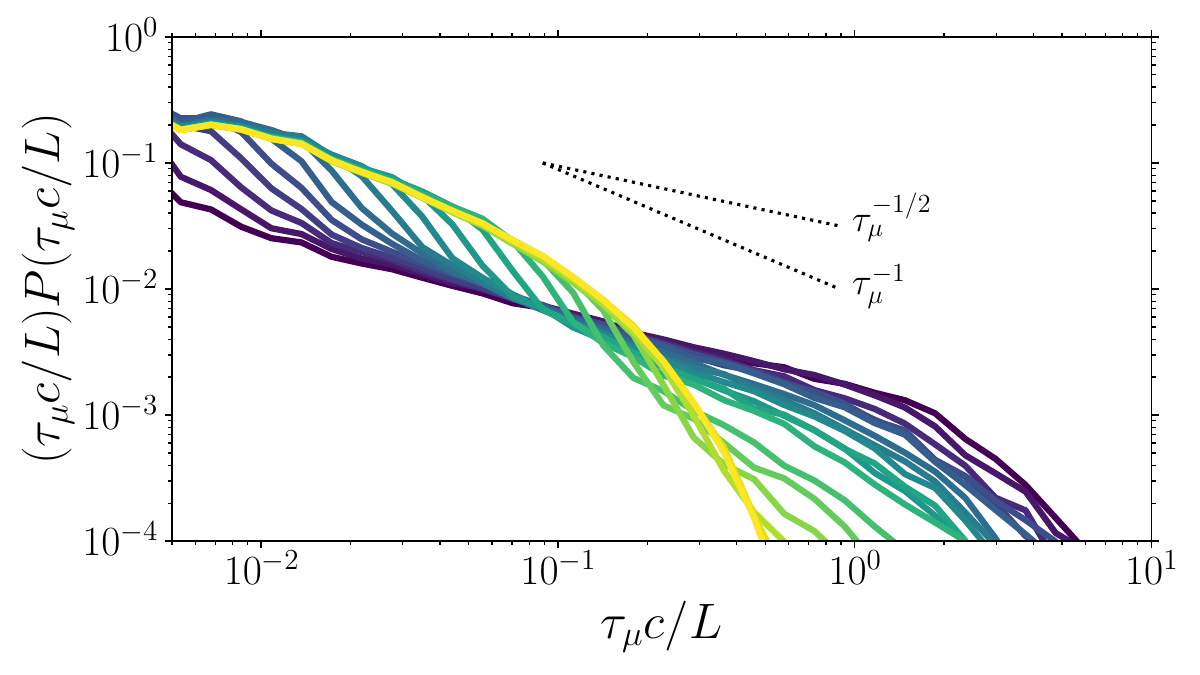}
        \end{minipage}   

  \caption{PDFs of particle collision times $P(\tau c/L)$ multiplied by $\tau c/L$ (thus showing the probability per logarithmic interval). We consider two types of spatial scattering events: (1) $\bb{v}\bcdot\eb$ changing sign (top panel) and (2) $\bb{v}\bcdot\ez$ changing sign (middle panel), and we define the associated collision times $\tau_b$ and $\tau_z$, respectively.  The locations of the peaks in the PDFs of $\tau_z$ are set by the gyro-periods. For small $r_L$, $P(\tau)$ shows clear power-law tails. For comparison we also plot as dotted lines the $\tau^{-1/2}$ Sparre-Andersen scaling and a $\tau^{-1}$ scaling. If the power-law is shallower then $\tau^{-1}$, as is the case over a wide range $r_L$, the global diffusion rate is set not by the mean free path but instead by the longest excursions (the location of the cutoff). In the top panel, the cutoff increases with decreasing $r_L$,  indicating that particles are able to avoid higher-curvature regions for increasingly long stretches of time. This is not the case in the middle panel, as the field-line random walk (dashed line; traced by particles with $\bb{v} =  c\eb$) imposes an upper limit on how far particles can propagate without changing their $v_z$. The different behavior of the two cutoffs is the reason why $\kappa_z$ and $\kappa_\parallel$ are different functions of $r_L$ (Figure~\ref{fig:diff_coeffs}). In the bottom panel, we show PDFs of collision times calculated using the particles' magnetic moments (cf. Figure~\ref{fig:mu_track}). We bin $\log(\mu_{\rm M})$ in bins of width $1/2$ and define $\tau_\mu$ as the time a particle spends in the same bin. $P(\tau_\mu)$ also shows similar power-law behavior. \label{fig:pdf_tau}}
\end{figure}

Figure~\ref{fig:mu_track} suggests that there is not a single characteristic time interval between individual scattering events. Instead, scattering times appear to be sampled from a broad distribution (cf. \citealt{Liang2025}). This conclusion is validated by Figure~\ref{fig:pdf_tau}, in which we show the distribution of scattering times $P(\tau c/L)$ obtained from tracks of $2048$ magnetic-field lines and $2048$ particles in each of the 18~mass bins, tracked for 10~box light-crossing times. Since we use $c=1$ and $L=1$, we will replace $P(\tau c/L)$ with just $P(\tau)$ for the remainder of this work.  We define two types of spatial scattering events: (1) $\bb{v}\bcdot\eb$ changing sign and (2) $\bb{v}\bcdot\ez$ changing sign. For type-1 events we define the collision time $\tau_b$; for type-2 events we define the collision time $\tau_z$. We also consider the distribution of collisions that change the particles' magnetic moment (Figure~\ref{fig:mu_track}). In particular, we construct bins in $\log(\mu_{\rm M})$ of width $1/2$ and define $\tau_\mu$ as the time it takes for a particle in a given bin to change its magnetic moment enough to move to a different bin. We show the distributions of $\tau_b$, $\tau_z$ and $\tau_\mu$ in the top, middle and bottom panels of Figure~\ref{fig:pdf_tau}, respectively.

It is immediately apparent from Figure~\ref{fig:pdf_tau} that the distributions of collision times are indeed broad and for small $r_L$  show clear power-law tails, $P(\tau) \sim \tau^{-\alpha}$ and $\alpha \in [-1.5, -2]$. Interestingly, the fact that $\alpha$ is close to $-1.5$, especially for the smallest $r_L$, strongly resembles the Sparre--Andersen scaling of first-passage times $t$, $P(t) \sim t^{-3/2}$, i.e., the distribution of times taken for particles diffusing symmetrically in 1D to first cross a point that is some distance from the particles' starting location (\citealt{ANDERSEN1954}; see also, \citealt{Redner_2001}; \citealt{Palyulin2019}; \citealt{Liang2025}). In particular, when applied to scattering in velocity space, the Sparre--Andersen scaling suggests a PDF of collision times $P(\tau) \sim \tau^{-3/2}$. However, we note that many of the assumptions underlying the Sparre--Andersen scaling are not valid in our case. For example, the background guide field introduces an asymmetry and renders the problem not truly Markovian. Trapping in scattering structures that persist  in time and result in correlated scattering also violates the memory-free assumption of the Sparre--Andersen scaling. Thus, deviations from the $\tau^{-3/2}$ scaling are generally expected.  

The locations of the peaks in the PDFs of $\tau_z$ (middle panel) are set by the gyro-period and thus depend on particle energy via $\gamma$ (if a field line is tilted with respect to the $z$-direction, then as a particle gyrates, its $v_z$ undergoes oscillations). The gyro-period also determines the minimum $\tau$ where the power-law behavior starts. We note that the power-law slopes of the $\tau_b$ (top panel) and $\tau_z$ (middle panel) distributions are quite similar. However, while the cutoff at large $\tau_b$ in the top panel moves to the right as the gyro-radius is decreased, the trend is opposite in the $\tau_z$ PDF in the middle panel. What the top panel shows physically is that particles are capable of avoiding resonant-curvature regions for increasingly long stretches of time as the curvature required for resonance becomes very large. To understand why this behavior is not present in the middle panel, it is useful to consider the distribution of ``scattering times" $\tau_z$ of the field lines as traced by particles with velocities $\bb{v} = c \eb$, i.e., the time intervals between events of $c\eb\bcdot\ez$ changing sign. This PDF is shown as the black dashed line. Interestingly, the field lines also show a cutoff in $\tau_z$, which means that in $\delta B_{\rm rms} / B_0 \approx 2$ turbulence, almost all field lines, when integrated for long enough ($\gtrsim$~one box light-crossing time), reverse direction with respect to the direction of the guide field. This cutoff in the field-line PDF is important as it also sets a natural cutoff in $\tau_z$ for CR particles executing long flights along the local magnetic-field direction: beyond a certain distance (time) a particle's $v_z$ changes sign not necessarily because it is scattered, but because the field line to which it is tied reverses direction. This is the reason why the PDF of $\tau_z$ has a relatively constant cutoff for $r_L/L\lesssim 10^{-2}$, i.e., for particles that are on average well magnetized in the simulation. Particles with larger $r_L$ are not as well confined by the field lines and are more likely to drift away from them as the field lines bend and/or fold onto themselves  (as seen over a wider range of particle energies in \citealt{Kempski2023}). As a result, for larger $r_L$ the cutoff in $\tau_z$ increases appreciably, which in turn results in $\kappa_z$ showing stronger energy dependence in that energy range (Figure~\ref{fig:diff_coeffs}), as we explain below. We note that the different normalization of the power-law tails in $P(\tau_z)$ is due to the fact that lower-energy particles ``collide" with respect to the $z$-direction also because of their gyration (the peak of the PDF), which occurs more frequently at smaller energies. From the bottom panel, we see that $P(\tau_\mu)$ also shows the characteristic power-law behavior, suggesting that the power-law in the distributions of $\tau_z$ and $\tau_b$ is not (only) because of magnetic mirroring, which conserves $\mu_{\rm M}$. Finally, although not shown here explicitly, we note that the PDFs of $\tau_x$ and $\tau_y$, defined analogously to $\tau_z$, are qualitatively similar to the PDF of $\tau_z$, except that the cutoff occurs at smaller times. 

The fact that the cutoff in $\tau_b$ increases with decreasing $r_L$, while the cutoff in $\tau_z$ is relatively constant, helps us understand the quite different energy dependence of $\kappa_\parallel$ compared to diffusion along globally specified axes, viz.~$\kappa_z$ and $\kappa_{xy}$ in Figure~\ref{fig:diff_coeffs}. First, we note that the slope of the PDF for small $r_L$ is quite shallow, $\alpha < 2$, and implies that the largest excursions (i.e., the location of the power-law cutoff) set global diffusion properties. Indeed, for an arbitrary distribution of free paths $P(\lambda)$ that has a cutoff, the diffusion coefficient is given by (\citealt{Bouchaud1990})
\begin{equation} \label{eq:kappa_arbitrary_pdf}
    \kappa = \frac{\langle  \lambda^2 \rangle}{2 \langle \tau \rangle} \sim \frac{1}{2} c^2 \frac{\langle  \tau^2 \rangle}{\langle \tau \rangle},
\end{equation}
where in the last step we used the rough relation that $\lambda \sim c \tau$. Now consider a power-law PDF  that has the form $P(\tau)=(1-\alpha)\tau^{-\alpha}\tau_{\rm min}^{\alpha-1}$, $1<\alpha<2$, in the range $\tau_{\rm min} < \tau < \tau_{\rm max} $ with $\tau_{\rm max} \gg \tau_{\rm min}$. In this case, we find that,
\begin{equation} 
    \kappa \sim \frac{1}{2} c^2 \frac{(1-\alpha) \tau_{\rm min}^{\alpha-1} \int_{\tau_{\rm min}}^{\tau_{\rm max}} \tau^{-\alpha+2} \rmd\tau}{(1-\alpha) \tau_{\rm min}^{\alpha-1} \int_{\tau_{\rm min}}^{\tau_{\rm max}} \tau^{-\alpha+1} \rmd\tau } \sim \frac{2-\alpha}{6-2\alpha}c^2 \tau_{\rm max}.
    \label{eq:kappa_taumax}
\end{equation}
Thus, assuming that the power-law exists over a sufficiently large range of scales, the overall diffusion is solely set by the largest excursions. Because our simulations show that $\tau_{z, \rm max}$ is a weak function of $r_L$, we find a very weak dependence on energy in $\kappa_z$. Moreover, using that $\alpha \approx 5/3$ is reasonable approximation for a range of $r_L$ and that $\tau_{z, \rm max}\approx L/c$ according to Figure~\ref{fig:pdf_tau}, equation~\eqref{eq:kappa_taumax} predicts that $\kappa_z \approx 0.1 cL$, which is roughly consistent with our diffusion results (Figure~\ref{fig:diff_coeffs}).  In contrast, $\tau_{b,\rm max}$ increases more appreciably with decreasing $r_L$, and so we find that $\kappa_\parallel$ increases with decreasing $r_L$.   

The fact that the rate of CR diffusion at small $r_L$ is dominated by the largest excursions, which along the cardinal directions are limited by the field-line random walk, is likely the reason why $\kappa_{xy}$ is so remarkably converged with resolution (orange curves in Figure~\ref{fig:diff_coeffs}). In particular, we find that $\tau_{\rm max}$ of field lines does not show any significant dependence on resolution and is therefore set by the large scales. Equation~\eqref{eq:kappa_taumax} then suggests that the rate of diffusion of small-$r_L$ particles should depend only weakly on resolution. However, this is only true in the $x$-$y$ plane and not along the $z$-direction, as demonstrated by Figure~\ref{fig:diff_coeffs}. The lack of convergence along $z$ is due to the presence of the background guide field and due to the fact that the lowest-energy particles are able to propagate very large distances along field lines without scattering (top panel of Figure~\ref{fig:pdf_tau}). In particular, if $  v_B \tau_b \gtrsim L$, particles sample a large enough volume to experience long-range correlations in the magnetic-field direction due to the background guide field, which reduces the effectiveness with which the field-line random walk (which along $z$ is a biased random walk) can inhibit their diffusion along $z$. This enhances diffusion along $z$ and renders it more similar to diffusion along the local magnetic field, such that $\kappa_z$ and $\kappa_\parallel$ roughly satisfy \eqref{eq:kappaz_kappapar} (Figure~\ref{fig:diff_coeffs}). This is why $\kappa_z$ starts increasing with decreasing $r_L$ at the smallest $r_L$, even though there is no apparent increase in the cutoff in the PDF of $\tau_z$ at the corresponding $r_L$ (Figure~\ref{fig:pdf_tau}). This increase in $\kappa_z$ at the smallest $r_L$ occurs at all resolutions investigated, but the critical value of $r_L$ decreases with increasing resolution. This is because pitch-angle scattering of small-gyro-radius particles along field lines becomes more efficient as resolution is increased and more high-curvature structures are resolved. The larger scattering rate of small-rigidity particles along field lines at the highest resolution is thus essential for ensuring that the field-line random walk can inhibit the diffusion along $z$ for a wide range of particle energies.

We stress that, while the largest excursions set the global diffusion rate, the scattering times form a self-similar L\'evy distribution and most scattering events occur on timescales $\tau \ll \tau_{\rm max}$. The nature of CR scattering in intermittent MHD turbulence is thus fundamentally different from the standard picture in which particle transport can be described solely in terms of a characteristic mean free path. 

\section{Discussion} \label{sec:discussion}
\subsection{Relationship to CR observations}
Recent observations by AMS-02 (\citealt{aguilar_2015}; \citealt{Aguilar2015_He}), DAMPE (\citealt{DAMPE2019}; \citealt{DAMPE2024}) and CALET (\citealt{Adriani2019}; \citealt{Adriani2023}) reveal significant spectral hardening in CR primary spectra (H, He, etc.) above a few hundred GeV, with a change in spectral index of order $\approx 0.2$ (\citealt{AGUILAR2021}). At the same time, measurements by AMS-02 (\citealt{AGUILAR2021}), DAMPE (\citealt{DAMPE2022}) and  CALET (\citealt{Adriani2022}) reveal a flattening in the CR boron-to-carbon (B/C) ratio  around a $\approx$TeV/nucleon. The B/C ratio is commonly used as a direct probe of CR transport in the Galaxy and the flattening suggests that there may be significant weakening in the observed energy dependence of CR transport. This has motivated a number of recent models that consider energy-independent transport above a few hundred GeV (\citealt{Pezzi2024}; \citealt{Recchia2024}). In addition, measurements of the anisotropy of CR arrival directions also show very weak energy dependence above a TeV (see \citealt{Burch2010} or \citealt{Blasi_Amato2012} for a compilation of measurements). This is most easily explained if the transport of CRs has weak dependence on energy in that range, although as shown in \cite{Blasi_Amato2012} this may also be explained by considering different distributions of CR sources.

The efficient scattering that we measure in our simulation ($\kappa_z \ll cL$), combined with its weak energy dependence, may offer a relatively simple explanation to this broad range of observations of CRs with energies greater than a few hundred GeV. The observed B/C ratio appears to flatten at a value of order ${\rm B/C(TeV/n)}\approx 0.04$. Assuming that the spallation reactions that produce boron nuclei occur primarily in a thin disk of half-thickness $h$ characterized by a spallation timescale $\tau_s$, and that the CRs diffuse with diffusion coefficient $\kappa$ over a CR halo size $H$ before escaping into the outer CGM, the predicted B/C $
\sim hH/(\tau_s \kappa)$. Assuming $h \approx 100$~pc, a thin-disk mean density $n\approx 1 \  {\rm cm}^{-3}$ and a spallation cross section $\sigma_s \approx 60$~mB, we find that the observed ${\rm B/C} \approx 0.04$ corresponds to an isotropic diffusion coefficient of order $5 \times 10^{29} \ {\rm cm^2 \ s^{-1}}$. In our highest-resolution simulation we find a minimum $\kappa_z \approx 0.07 cL$. If, for simplicity, we assume that the entire Galaxy is filled with mildly super-Alfv\'enic turbulence as in our simulation with characteristic driving scale $L \approx 100$~pc, we find $\kappa_z \approx 7 \times 10^{29}\ {\rm cm^2 \ s^{-1}}$. This is comparable to the value inferred from the observed B/C flattening, although we note that our assumption of uniformly driven turbulence in the entire volume of the Galaxy is hardly realistic. Moreover, Figure~\ref{fig:diff_coeffs} clearly shows that assuming an isotropic diffusion coefficient is not well motivated, as we find that $\kappa_{xy}\ll \kappa_z$. Nevertheless, it is reassuring that our numerical results are at first glance not incompatible with observations.

Below a few hundred GeV, the observed energy dependence of transport is stronger (\citealt{aguilar_bc}) and appears to follow the scaling $\kappa \sim E^\delta$ with $\delta \sim 0.3-0.6$.\footnote{ We note, however, that energy-independent diffusion was also recently inferred for $\sim$GeV CR electrons in the star-forming environments of M51 by \cite{Heesen2023}.} This strong scaling is not present in our simulations, and so we now discuss what additional physics may be needed to explain it. 

\subsection{PDF of collision times and energy dependence}
The dependence of $\kappa_z$ on particle energy is weak in our simulations because the largest excursions set the diffusion rate, and these excursions are in turn limited by field lines changing direction. The largest excursions dominate because the power-law slope of the PDF of collision times is sufficiently shallow. More generally, $\tau_{\rm max}$ setting $\kappa$ as in \eqref{eq:kappa_taumax} holds for $P(\tau) \sim \tau^{-\alpha}$ and $1<\alpha<2$. If $2<\alpha<3$, we instead find that
\begin{equation} \label{eq:kappa_taumax_steeper}
    \kappa \propto  \tau_{\rm max}^{3-\alpha} \tau_{\rm min}^{-2+\alpha} \propto \tau_{\rm max}^{3-\alpha} r_L^{-2 + \alpha},
\end{equation}
where in the last step we used that $\tau_{\rm min}$ is related to the particle gyro-period. The dependence on $\tau_{\rm min}$ that we now get is of possible importance. In particular, for $\alpha=7/3$ and fixed $\tau_{\rm max}$, this  would produce the scaling $\kappa \sim r_L^{1/3}$, which matches the scaling supported by measurements of CRs in the Milky Way below a few hundred GeV (\citealt{aguilar_bc}; \citealt{amato_blasi_18}).

The slopes of $P(\tau)$ and their origins are thus critical for understanding the energy dependence of transport. The slopes of $P(\tau_{\mu})$ and $P(\tau_b)$ likely depend on the clustering properties of resonant high-curvature regions and how long particles are trapped inside them. This is beyond the scope of this Letter and will be explored in more detail in subsequent work. However, we point out that a power-law distribution of collision times may be a general property of CR transport. This could be due to the scaling laws associated with idealized diffusion processes, such as the Sparre--Andersen scaling (\citealt{ANDERSEN1954}; \citealt{Palyulin2019}), which when applied to scattering in velocity space result in a power-law distribution of collision times, and may be further driven by the self-similarity of turbulence, in particular the statistical properties of large-amplitude MHD turbulence characterized by power-law PDFs of magnetic field-line curvature. To illustrate the latter, it is useful to consider a toy model in which regions characterized by different values of curvature $K$ act as distinct families of objects that particles can scatter off: if the curvature is resonant, the particles are scattered while changing their magnetic moment, and if $K$ is small relative to the inverse gyro-radius, then the particles are scattered in the spatial sense because they adiabatically follow a bending field line. Crucially, with respect to the cardinal axes, particles are scattered by a wide range of curvatures as long as $K\lesssim r_L^{-1}$, as discussed in \cite{Kempski2023}. Next, suppose that regions of curvature $K$ introduce their own mean free path, $\lambda(K)$,  via,\footnote{Here we are ignoring any clustering and that $P(\lambda)$ is likely a broad power-law PDF even at fixed $K$.}
\begin{equation} \label{eq:lambda_K}
    \lambda(K) \sim \frac{1}{n(K) \sigma(K)} \sim \frac{1}{P(K) K} ,
\end{equation}
where $\sigma(K)$ is the typical cross section of a region of curvature $K$ and $n(K)$ is the number density of such regions. In the last step we replaced $n(K)$ by the volume filling fraction via $n(K) \sim P(K) / [K^{-1} \sigma (K)]$ and we additionally assumed that a curvature region $K$ typically has a short-axis length of order $K^{-1}$. For $P(K)\sim K^{-\beta}$, with $\beta \approx 2.5$ in $\delta B_{\rm rms}/B_0 \gtrsim 1$ turbulence \citep{Yang2019, Bandyopadhyay2020, Lemoine2023, Kempski2023}, equation~\eqref{eq:lambda_K} implies that $\lambda(K) \sim K^{\beta-1}$. Thus, the mean free path introduced by each curvature is typically an increasing function of $K$. Combining the free paths from all $K$, we get the PDF of $\lambda$,
\begin{equation}
    P(\lambda) = P(K) \frac{\rmd K}{\rmd \lambda} \sim K^{-\beta} \lambda^{\frac{2-\beta}{\beta-1}} \sim \lambda^{-2}. 
\end{equation}
Thus, a power-law PDF of field-line reversals (without invoking any clustering) suggests a power-law distribution of free paths along the cardinal axes. If we also use that the collision cross section $\sigma(K)$ is effectively smaller due to magnetic defocusing with $\sigma_{\rm eff}(K) \sim \sigma(K) B(K) / B_{\rm rms} \sim \sigma(K) K^{-1/2}$ (since high-curvature regions are characterized by weaker magnetic fields, in particular $\langle B(K) \rangle \sim K^{-1/2}$; \citealt{schekochihin_2004}), then we instead find that  the slope of $P(\lambda)$ depends on $\beta$ and for $\beta=2.5$ we get $P(\lambda) \sim \lambda^{-7/4}$. We stress that the above derivations are not meant to represent theoretical models of the measured PDFs since we are neglecting the clustering of scattering sites and/or the trapping of particles. Instead, our goal is to illustrate that power-law distributions of free paths may be a general property of transport in large-amplitude turbulence characterized by field reversals on all scales.

As shown in the middle panel of Figure~\ref{fig:pdf_tau}, the $P(\tau_z)$ of particles exhibits a steeper slope than the field lines. This is likely because field lines are better at avoiding high-curvature regions associated with weak magnetic fields (i.e., magnetic defocusing is generally more effective for field lines than for particles whose gyro-radii grow in weaker magnetic fields) and due to the fact that particles can repeatedly bounce between scattering sites, unlike field lines. Thus, a mechanism that promotes more efficient trapping of particles between scattering sites appears to be a natural path toward a steeper $P(\tau)$ and a stronger energy dependence of transport. This could, for example, occur if the clustering properties of high-curvature regions are modified by a well-resolved distribution of plasmoid sizes inside reconnection layers \citep{Huang2012, Petropoulou2018,Majeski2021}. Alternatives include more laminar field lines between scattering sites to slow down the untrapping process, or more efficient trapping between magnetic mirrors (e.g., \citealt{xu_2020_trapping}; \citealt{lazarian_xu_mirror}) than in our simulations. These properties likely depend on the presence of additional physical processes that go beyond the ideal-MHD and test-particle assumptions used here.

An important ingredient that is missing from our simulation is the dynamical feedback from CRs with energies less than a few hundred GeV, which are energetically important and can change the turbulence by depositing significant amounts of energy and momentum, and exciting instabilities (\citealt{kp69}; \citealt{zweibel_micro}). The feedback from CRs is likely important for explaining the observed stronger energy dependence of CR transport at low energies (e.g., \citealt{blasi12}; \citealt{Armillotta2025}).

\subsection{Relation to previous work} 
As in \cite{Lemoine2023} and \cite{Kempski2023}, we find that regions of high magnetic-field-line curvature are key mediators of particle transport. However, the energy dependence of transport that we measure here is significantly different from the one measured in \cite{Kempski2023} or predicted in \cite{Lemoine2023}. \cite{Lemoine2023} predicted an energy-dependent mean free path with scaling $\ell \propto r_L^{0.3}$ based on PDFs of coarsened field-line curvature. The key difference in our work is the measured broad distribution of collision times, which implies that the overall transport rate is determined by the largest excursions rather than the mean free path. This results in a significantly weaker energy dependence. A stronger energy dependence down to small gyro-radii ($r_L/L \sim 10^{-3}$) was also measured in \cite{Kempski2023}, which was interpreted as being the product of varying degrees of magnetization of particles as they propagated in magnetic folds of different sizes. However, the limited resolution and the larger amplitude of the turbulence simulations in \cite{Kempski2023} resulted in significantly smaller Alfv\'en scales and as a result the turbulence was characterized by volume-filling perpendicular field reversals close to the dissipation scale. This resulted in strong energy dependence. In this work, the overall transport has a much weaker energy dependence (in addition to being significantly more anisotropic). We see strong energy dependence in $\kappa_z$ only for the largest gyro-radii, $r_L/L \gtrsim 4 \times 10^{-2}$, i.e., gyro-radii exceeding the Alfv\'en scale, followed by weak energy dependence over a very wide range in energy. This underscores the importance of resolving the Alfv\'en scale in large-amplitude MHD turbulence. \cite{Kempski2023} reported weak energy dependence only at the smallest $r_L$ (which were not properly resolved) and argued that the weak energy dependence was due to the field-line random walk setting the transport rate. Interestingly, here we find that the field-line random walk remains important even in the presence of a significant mean field, and determines the transport rate over a wide range of resolved particle energies (by limiting the particles' largest excursions). Our follow-up work, which will explore a range of $\delta B_{\rm rms}/B_0$, will be well suited to further connect the present findings with those of \cite{Kempski2023}.

\section{Summary} \label{sec:summary}
In this work, we have studied the propagation of charged cosmic rays (CRs) through a static snapshot of MHD turbulence simulated at extremely high resolution equal to $10{,}240^3$ grid cells (Figure \ref{fig:j_particle}; Fielding et al., in prep.). Even though the turbulence is super-Alfv\'enic ($\delta B_{\rm rms}/B_0=2$), we find that the transport is highly anisotropic with respect to the direction of the background guide field (Figure~\ref{fig:diff_coeffs}). The CRs in our simulation are scattered efficiently via \textit{resonant-curvature} scattering (Figures~\ref{fig:diff_coeffs} and \ref{fig:mu_track}), as previously proposed by \cite{Lemoine2023} and \cite{Kempski2023} based on simulations either without background guide field or a significantly weaker guide field. In follow-up work, we use a variety of initial conditions at slightly lower resolution to study particle transport in turbulence characterized by a range of $\delta B_{\rm rms}/B_0$. Neither here nor in the broader range of simulations considered in our follow-up study do we find any evidence for significant CR scattering by small-amplitude waves (Kempski et al., in prep.),\footnote{We employ fully solenoidal driving here, which is not effective at generating fast modes. However, in \cite{Kempski2023} we found that $50$\% compressible driving did not change CR transport.} one of the prevailing models for CR transport for much of the last ${\sim}50$ years.

At a resolution of $10{,}240^3$, we start to see an extended range of CR energies at which the rate of diffusion has a weak dependence on CR energy (Figure~\ref{fig:diff_coeffs}). At these energies, particle scattering is characterized by a power-law PDF of scattering times (Figure~\ref{fig:pdf_tau}), suggesting that we are capturing a self-similar regime of particle transport. We find that the rate of particle diffusion is set by the particles' largest free excursions, which are very rare but nevertheless dominate the transport. The largest excursions are limited by the random walk of the turbulent magnetic-field lines; this is why the overall rate of transport has a weak dependence on particle energy. 

Our results support the idea that regions of large magnetic-field-line curvature are efficient scatterers of CRs even on small scales at which the median $\delta B (\ell) \ll B_0$, and that CR transport in galaxies may indeed be highly intermittent (as hypothesized by, e.g.,  \citealt{Butsky2024}). In addition, our analysis suggests that CR transport is not well described by a simple diffusion process that can be fully described by a characteristic mean free path, and therefore more careful treatments that incorporate fractionally diffusive processes in galactic propagation are likely necessary (e.g., as recently studied by \citealt{Liang2025}).

\section*{Data availability}
The calculations from this article will be shared on reasonable request to the corresponding author.

\section*{Acknowledgements}
We thank Lucia Armillotta, Iryna Butsky, Luca Comisso, Minghao Guo, Philip Hopkins, Rajsekhar Mohapatra, Siang Peng Oh, Eve Ostriker, Sarah Recchia, Lorenzo Sironi, Anatoly Spitkovsky, Jonathan Squire, Mila Winter and Ellen Zweibel for useful discussions. The simulations performed in this work used resources of the Oak Ridge Leadership Computing Facility, which is a DOE Office of Science User Facility supported under Contract DE-AC05-00OR22725. These computational resources were provided as part of the DOE INCITE Leadership Computing Program under allocation AST-207 (PI: Fielding). We thank the Oak Ridge Leadership Computing Facility support staff, in particular Vassilios Mewes, as well as Nrushad Joshi, for their continuous support throughout the project. This research was supported in part by grant NSF PHY-2309135 to the Kavli Institute for Theoretical Physics (KITP), and performed in part at Aspen Center for Physics, which is supported by National Science Foundation grant PHY-2210452. PK was supported by the Lyman Spitzer, Jr. Fellowship at Princeton University. DBF was supported in part by NSF AAG Award No.~2407387. MWK was supported in part by NSF CAREER Award No.~1944972. RJE and AP acknowledge support from the Simons Foundation (MP-SCMPS-00001470). AP was partially supported by an Alfred P. Sloan Research Fellowship and a Packard Foundation Fellowship in Science and Engineering. This research was facilitated by the Multimessenger Plasma Physics Center (MPPC), NSF grants PHY-2206607 and PHY-2206610 (EQ and AP).



\appendix

\section{PDF of Field-Line Curvature}
\begin{figure}
    \centering
    \includegraphics[width=0.5\linewidth]{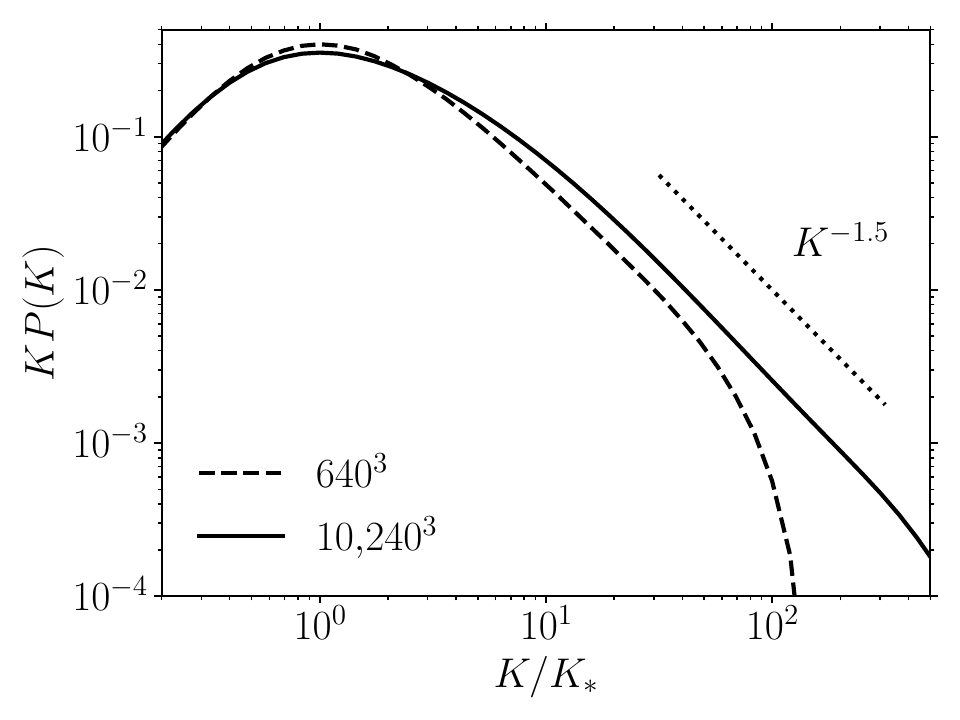}
    \caption{PDF of field-line curvature $K$ normalized by the curvature $K_*$ that maximizes $KP(K)$ (i.e. the most volume-filling logarithmic curvature bin) for the $640^3$ and $10{,}240^3$ simulations. Both PDFs are characterized by a $K^{-1.5}$ power-law tail at high $K$. However, as resolution is increased, the width of the PDF near the peak is broadened significantly. At $10{,}240^3$, the power-law occurs for $K\gtrsim10K_*$, which is significantly larger than the corresponding threshold in the $640^3$ simulation. We speculate that it is this broadening of the PDF that is responsible for the more efficient scattering along field lines (pink lines in Figure \ref{fig:diff_coeffs}) at higher resolution. See Fielding et al. (in prep.) for more details.   }
    \label{fig:pdf_K_resolution}
\end{figure}

In Figure \ref{fig:pdf_K_resolution}, we show the PDFs of field-line curvature, $K$, in the $640^3$ and $10{,}240^3$ simulations, normalized by the curvature $K_*$ that maximizes $KP(K)$, i.e., the probability per logarithmic curvature interval. Both PDFs approach the characteristic $K^{-1.5}$ power-law tail at high $K$ \citep{Yang2019, Bandyopadhyay2020}. However, in the higher-resolution simulation, the width of the PDF near the peak is significantly larger.

\bibliography{test_particle}{}

\begin{thebibliography}{}
\expandafter\ifx\csname natexlab\endcsname\relax\def\natexlab#1{#1}\fi
\providecommand{\url}[1]{\href{#1}{#1}}
\providecommand{\dodoi}[1]{doi:~\href{http://doi.org/#1}{\nolinkurl{#1}}}
\providecommand{\doeprint}[1]{\href{http://ascl.net/#1}{\nolinkurl{http://ascl.net/#1}}}
\providecommand{\doarXiv}[1]{\href{https://arxiv.org/abs/#1}{\nolinkurl{https://arxiv.org/abs/#1}}}

\bibitem[{{Adriani} {et~al.}(2019){Adriani}, {Akaike}, {Asano}, {Asaoka}, {Bagliesi}, {Berti}, {Bigongiari}, {Binns}, {Bonechi}, {Bongi}, {Brogi}, {Bruno}, {Buckley}, {Cannady}, {Castellini}, {Checchia}, {Cherry}, {Collazuol}, {di Felice}, {Ebisawa}, {Fuke}, {Guzik}, {Hams}, {Hasebe}, {Hibino}, {Ichimura}, {Ioka}, {Ishizaki}, {Israel}, {Kasahara}, {Kataoka}, {Kataoka}, {Katayose}, {Kato}, {Kawanaka}, {Kawakubo}, {Kohri}, {Krawczynski}, {Krizmanic}, {Lomtadze}, {Maestro}, {Marrocchesi}, {Messineo}, {Mitchell}, {Miyake}, {Moiseev}, {Mori}, {Mori}, {Mori}, {Motz}, {Munakata}, {Murakami}, {Nakahira}, {Nishimura}, {de Nolfo}, {Okuno}, {Ormes}, {Ozawa}, {Pacini}, {Palma}, {Papini}, {Penacchioni}, {Rauch}, {Ricciarini}, {Sakai}, {Sakamoto}, {Sasaki}, {Shimizu}, {Shiomi}, {Sparvoli}, {Spillantini}, {Stolzi}, {Suh}, {Sulaj}, {Takahashi}, {Takayanagi}, {Takita}, {Tamura}, {Terasawa}, {Tomida}, {Torii}, {Tsunesada}, {Uchihori}, {Ueno}, {Vannuccini}, {Wefel}, {Yamaoka}, {Yanagita}, {Yoshida}, {Yoshida}, \& {Calet
  Collaboration}}]{Adriani2019}
{Adriani}, O., {Akaike}, Y., {Asano}, K., {et~al.} 2019, \prl, 122, 181102, \dodoi{10.1103/PhysRevLett.122.181102}

\bibitem[{{Adriani} {et~al.}(2022){Adriani}, {Akaike}, {Asano}, {Asaoka}, {Berti}, {Bigongiari}, {Binns}, {Bongi}, {Brogi}, {Bruno}, {Buckley}, {Cannady}, {Castellini}, {Checchia}, {Cherry}, {Collazuol}, {de Nolfo}, {Ebisawa}, {Ficklin}, {Fuke}, {Gonzi}, {Guzik}, {Hams}, {Hibino}, {Ichimura}, {Ioka}, {Ishizaki}, {Israel}, {Kasahara}, {Kataoka}, {Kataoka}, {Katayose}, {Kato}, {Kawanaka}, {Kawakubo}, {Kobayashi}, {Kohri}, {Krawczynski}, {Krizmanic}, {Maestro}, {Marrocchesi}, {Messineo}, {Mitchell}, {Miyake}, {Moiseev}, {Mori}, {Mori}, {Motz}, {Munakata}, {Nakahira}, {Nishimura}, {Okuno}, {Ormes}, {Ozawa}, {Pacini}, {Papini}, {Rauch}, {Ricciarini}, {Sakai}, {Sakamoto}, {Sasaki}, {Shimizu}, {Shiomi}, {Spillantini}, {Stolzi}, {Sugita}, {Sulaj}, {Takita}, {Tamura}, {Terasawa}, {Torii}, {Tsunesada}, {Uchihori}, {Vannuccini}, {Wefel}, {Yamaoka}, {Yanagita}, {Yoshida}, {Yoshida}, {Zober}, \& {Calet Collaboration}}]{Adriani2022}
---. 2022, \prl, 129, 251103, \dodoi{10.1103/PhysRevLett.129.251103}

\bibitem[{{Adriani} {et~al.}(2023){Adriani}, {Akaike}, {Asano}, {Asaoka}, {Berti}, {Bigongiari}, {Binns}, {Bongi}, {Brogi}, {Bruno}, {Buckley}, {Cannady}, {Castellini}, {Checchia}, {Cherry}, {Collazuol}, {de Nolfo}, {Ebisawa}, {Ficklin}, {Fuke}, {Gonzi}, {Guzik}, {Hams}, {Hibino}, {Ichimura}, {Ioka}, {Ishizaki}, {Israel}, {Kasahara}, {Kataoka}, {Kataoka}, {Katayose}, {Kato}, {Kawanaka}, {Kawakubo}, {Kobayashi}, {Kohri}, {Krawczynski}, {Krizmanic}, {Maestro}, {Marrocchesi}, {Messineo}, {Mitchell}, {Miyake}, {Moiseev}, {Mori}, {Mori}, {Motz}, {Munakata}, {Nakahira}, {Nishimura}, {Okuno}, {Ormes}, {Ozawa}, {Pacini}, {Papini}, {Rauch}, {Ricciarini}, {Sakai}, {Sakamoto}, {Sasaki}, {Shimizu}, {Shiomi}, {Spillantini}, {Stolzi}, {Sugita}, {Sulaj}, {Takita}, {Tamura}, {Terasawa}, {Torii}, {Tsunesada}, {Uchihori}, {Vannuccini}, {Wefel}, {Yamaoka}, {Yanagita}, {Yoshida}, {Yoshida}, {Zober}, \& {Calet Collaboration}}]{Adriani2023}
---. 2023, \prl, 130, 171002, \dodoi{10.1103/PhysRevLett.130.171002}

\bibitem[{{Aguilar} {et~al.}(2015{\natexlab{a}}){Aguilar}, {Aisa}, {Alpat}, {Alvino}, {Ambrosi}, {Andeen}, {Arruda}, {Attig}, {Azzarello}, {Bachlechner}, {Barao}, {Barrau}, {Barrin}, {Bartoloni}, {Basara}, {Battarbee}, {Battiston}, {Bazo}, {Becker}, {Behlmann}, {Beischer}, {Berdugo}, {Bertucci}, {Bigongiari}, {Bindi}, {Bizzaglia}, {Bizzarri}, {Boella}, {de Boer}, {Bollweg}, {Bonnivard}, {Borgia}, {Borsini}, {Boschini}, {Bourquin}, {Burger}, {Cadoux}, {Cai}, {Capell}, {Caroff}, {Casaus}, {Cascioli}, {Castellini}, {Cernuda}, {Cerreta}, {Cervelli}, {Chae}, {Chang}, {Chen}, {Chen}, {Cheng}, {Chen}, {Cheng}, {Chou}, {Choumilov}, {Choutko}, {Chung}, {Clark}, {Clavero}, {Coignet}, {Consolandi}, {Contin}, {Corti}, {Gil}, {Coste}, {Creus}, {Crispoltoni}, {Cui}, {Dai}, {Delgado}, {Della Torre}, {Demirk{\"o}z}, {Derome}, {Di Falco}, {Di Masso}, {Dimiccoli}, {D{\'\i}az}, {von Doetinchem}, {Donnini}, {Du}, {Duranti}, {D'Urso}, {Eline}, {Eppling}, {Eronen}, {Fan}, {Farnesini}, {Feng}, {Fiandrini}, {Fiasson}, {Finch},
  {Fisher}, {Galaktionov}, {Gallucci}, {Garc{\'\i}a}, {Garc{\'\i}a-L{\'o}pez}, {Gargiulo}, {Gast}, {Gebauer}, {Gervasi}, {Ghelfi}, {Gillard}, {Giovacchini}, {Goglov}, {Gong}, {Goy}, {Grabski}, {Grandi}, {Graziani}, {Guandalini}, {Guerri}, {Guo}, {Haas}, {Habiby}, {Haino}, {Han}, {He}, {Heil}, {Hoffman}, {Hsieh}, {Huang}, {Huh}, {Incagli}, {Ionica}, {Jang}, {Jinchi}, {Kanishev}, {Kim}, {Kim}, {Kirn}, {Kossakowski}, {Kounina}, {Kounine}, {Koutsenko}, {Krafczyk}, {La Vacca}, {Laudi}, {Laurenti}, {Lazzizzera}, {Lebedev}, {Lee}, {Lee}, {Leluc}, {Levi}, {Li}, {Li}, {Li}, {Li}, {Li}, {Li}, {Li}, {Li}, {Li}, {Lim}, {Lin}, {Lipari}, {Lippert}, {Liu}, {Liu}, {Lolli}, {Lomtadze}, {Lu}, {Lu}, {Lu}, {Luebelsmeyer}, {Luo}, {Lv}, {Majka}, {Ma{\~n}{\'a}}, {Mar{\'\i}n}, {Martin}, {Mart{\'\i}nez}, {Masi}, {Maurin}, {Menchaca-Rocha}, {Meng}, {Mo}, {Morescalchi}, {Mott}, {M{\"u}ller}, {Ni}, {Nikonov}, {Nozzoli}, {Nunes}, {Obermeier}, {Oliva}, {Orcinha}, {Palmonari}, {Palomares}, {Paniccia}, {Papi}, {Pauluzzi}, {Pedreschi},
  {Pensotti}, {Pereira}, {Picot-Clemente}, {Pilo}, {Piluso}, {Pizzolotto}, {Plyaskin}, {Pohl}, {Poireau}, {Postaci}, {Putze}, {Quadrani}, {Qi}, {Qin}, {Qu}, {R{\"a}ih{\"a}}, {Rancoita}, {Rapin}, {Ricol}, {Rodr{\'\i}guez}, {Rosier-Lees}, {Rozhkov}, {Rozza}, {Sagdeev}, {Sandweiss}, {Saouter}, {Sbarra}, {Schael}, {Schmidt}, {von Dratzig}, {Schwering}, {Scolieri}, {Seo}, {Shan}, {Shan}, {Shi}, {Shi}, {Shi}, {Siedenburg}, {Son}, {Spada}, {Spinella}, {Sun}, {Sun}, {Tacconi}, {Tang}, {Tang}, {Tang}, {Tao}, {Tescaro}, {Ting}, {Ting}, {Tomassetti}, {Torsti}, {T{\"u}rko{\v{g}}lu}, {Urban}, {Vagelli}, {Valente}, {Vannini}, {Valtonen}, {Vaurynovich}, {Vecchi}, {Velasco}, {Vialle}, {Vitale}, {Vitillo}, {Wang}, {Wang}, {Wang}, {Wang}, {Wang}, {Wang}, {Weng}, {Whitman}, {Wienkenh{\"o}ver}, {Wu}, {Wu}, {Xia}, {Xie}, {Xie}, {Xiong}, {Xin}, {Xu}, {Xu}, {Yan}, {Yang}, {Yang}, {Ye}, {Yi}, {Yu}, {Yu}, {Zeissler}, {Zhang}, {Zhang}, {Zhang}, {Zhang}, {Zheng}, {Zhuang}, {Zhukov}, {Zichichi}, {Zimmermann}, {Zuccon}, {Zurbach}, \&
  {AMS Collaboration}}]{aguilar_2015}
{Aguilar}, M., {Aisa}, D., {Alpat}, B., {et~al.} 2015{\natexlab{a}}, \prl, 114, 171103, \dodoi{10.1103/PhysRevLett.114.171103}

\bibitem[{{Aguilar} {et~al.}(2015{\natexlab{b}}){Aguilar}, {Aisa}, {Alpat}, {Alvino}, {Ambrosi}, {Andeen}, {Arruda}, {Attig}, {Azzarello}, {Bachlechner}, {Barao}, {Barrau}, {Barrin}, {Bartoloni}, {Basara}, {Battarbee}, {Battiston}, {Bazo}, {Becker}, {Behlmann}, {Beischer}, {Berdugo}, {Bertucci}, {Bindi}, {Bizzaglia}, {Bizzarri}, {Boella}, {de Boer}, {Bollweg}, {Bonnivard}, {Borgia}, {Borsini}, {Boschini}, {Bourquin}, {Burger}, {Cadoux}, {Cai}, {Capell}, {Caroff}, {Casaus}, {Castellini}, {Cernuda}, {Cerreta}, {Cervelli}, {Chae}, {Chang}, {Chen}, {Chen}, {Chen}, {Chen}, {Cheng}, {Chou}, {Choumilov}, {Choutko}, {Chung}, {Clark}, {Clavero}, {Coignet}, {Consolandi}, {Contin}, {Corti}, {Gil}, {Coste}, {Creus}, {Crispoltoni}, {Cui}, {Dai}, {Delgado}, {Della Torre}, {Demirk{\"o}z}, {Derome}, {Di Falco}, {Di Masso}, {Dimiccoli}, {D{\'\i}az}, {von Doetinchem}, {Donnini}, {Duranti}, {D'Urso}, {Egorov}, {Eline}, {Eppling}, {Eronen}, {Fan}, {Farnesini}, {Feng}, {Fiandrini}, {Fiasson}, {Finch}, {Fisher}, {Formato},
  {Galaktionov}, {Gallucci}, {Garc{\'\i}a}, {Garc{\'\i}a-L{\'o}pez}, {Gargiulo}, {Gast}, {Gebauer}, {Gervasi}, {Ghelfi}, {Giovacchini}, {Goglov}, {Gong}, {Goy}, {Grabski}, {Grandi}, {Graziani}, {Guandalini}, {Guerri}, {Guo}, {Haas}, {Habiby}, {Haino}, {Han}, {He}, {Heil}, {Hoffman}, {Hsieh}, {Huang}, {Huh}, {Incagli}, {Ionica}, {Jang}, {Jinchi}, {Kanishev}, {Kim}, {Kim}, {Kirn}, {Korkmaz}, {Kossakowski}, {Kounina}, {Kounine}, {Koutsenko}, {Krafczyk}, {La Vacca}, {Laudi}, {Laurenti}, {Lazzizzera}, {Lebedev}, {Lee}, {Lee}, {Leluc}, {Li}, {Li}, {Li}, {Li}, {Li}, {Li}, {Li}, {Li}, {Li}, {Li}, {Lim}, {Lin}, {Lipari}, {Lippert}, {Liu}, {Liu}, {Liu}, {Lolli}, {Lomtadze}, {Lu}, {Lu}, {Lu}, {Luebelsmeyer}, {Luo}, {Luo}, {Lv}, {Majka}, {Ma{\~n}{\'a}}, {Mar{\'\i}n}, {Martin}, {Mart{\'\i}nez}, {Masi}, {Maurin}, {Menchaca-Rocha}, {Meng}, {Mo}, {Morescalchi}, {Mott}, {M{\"u}ller}, {Nelson}, {Ni}, {Nikonov}, {Nozzoli}, {Nunes}, {Obermeier}, {Oliva}, {Orcinha}, {Palmonari}, {Palomares}, {Paniccia}, {Papi}, {Pauluzzi},
  {Pedreschi}, {Pensotti}, {Pereira}, {Picot-Clemente}, {Pilo}, \& {Piluso}}]{Aguilar2015_He}
---. 2015{\natexlab{b}}, \prl, 115, 211101, \dodoi{10.1103/PhysRevLett.115.211101}

\bibitem[{{Aguilar} {et~al.}(2016){Aguilar}, {Ali Cavasonza}, {Ambrosi}, {Arruda}, {Attig}, {Aupetit}, {Azzarello}, {Bachlechner}, {Barao}, {Barrau}, {Barrin}, {Bartoloni}, {Basara}, {Ba{\textcommabelow s}e{\v{g}}mez-du Pree}, {Battarbee}, {Battiston}, {Becker}, {Behlmann}, {Beischer}, {Berdugo}, {Bertucci}, {Bindel}, {Bindi}, {Boella}, {de Boer}, {Bollweg}, {Bonnivard}, {Borgia}, {Boschini}, {Bourquin}, {Bueno}, {Burger}, {Cadoux}, {Cai}, {Capell}, {Caroff}, {Casaus}, {Castellini}, {Cervelli}, {Chae}, {Chang}, {Chen}, {Chen}, {Chen}, {Cheng}, {Chou}, {Choumilov}, {Choutko}, {Chung}, {Clark}, {Clavero}, {Coignet}, {Consolandi}, {Contin}, {Corti}, {Creus}, {Crispoltoni}, {Cui}, {Dai}, {Delgado}, {Della Torre}, {Demakov}, {Demirk{\"o}z}, {Derome}, {Di Falco}, {Dimiccoli}, {D{\'\i}az}, {von Doetinchem}, {Dong}, {Donnini}, {Duranti}, {D'Urso}, {Egorov}, {Eline}, {Eronen}, {Feng}, {Fiandrini}, {Finch}, {Fisher}, {Formato}, {Galaktionov}, {Gallucci}, {Garc{\'\i}a}, {Garc{\'\i}a-L{\'o}pez}, {Gargiulo}, {Gast},
  {Gebauer}, {Gervasi}, {Ghelfi}, {Giovacchini}, {Goglov}, {G{\'o}mez-Coral}, {Gong}, {Goy}, {Grabski}, {Grandi}, {Graziani}, {Guo}, {Haino}, {Han}, {He}, {Heil}, {Hoffman}, {Hsieh}, {Huang}, {Huang}, {Huh}, {Incagli}, {Ionica}, {Jang}, {Jinchi}, {Kang}, {Kanishev}, {Kim}, {Kim}, {Kirn}, {Konak}, {Kounina}, {Kounine}, {Koutsenko}, {Krafczyk}, {La Vacca}, {Laudi}, {Laurenti}, {Lazzizzera}, {Lebedev}, {Lee}, {Lee}, {Leluc}, {Li}, {Li}, {Li}, {Li}, {Li}, {Li}, {Li}, {Li}, {Li}, {Lim}, {Lin}, {Lipari}, {Lippert}, {Liu}, {Liu}, {Lordello}, {Lu}, {Lu}, {Luebelsmeyer}, {Luo}, {Luo}, {Lv}, {Machate}, {Majka}, {Ma{\~n}{\'a}}, {Mar{\'\i}n}, {Martin}, {Mart{\'\i}nez}, {Masi}, {Maurin}, {Menchaca-Rocha}, {Meng}, {Mikuni}, {Mo}, {Morescalchi}, {Mott}, {Nelson}, {Ni}, {Nikonov}, {Nozzoli}, {Oliva}, {Orcinha}, {Palmonari}, {Palomares}, {Paniccia}, {Pauluzzi}, {Pensotti}, {Pereira}, {Picot-Clemente}, {Pilo}, {Pizzolotto}, {Plyaskin}, {Pohl}, {Poireau}, {Putze}, {Quadrani}, {Qi}, {Qin}, {Qu}, {R{\"a}ih{\"a}}, {Rancoita},
  {Rapin}, {Ricol}, {Rosier-Lees}, {Rozhkov}, {Rozza}, {Sagdeev}, {Sandweiss}, {Saouter}, {Schael}, {Schmidt}, {Schulz von Dratzig}, {Schwering}, {Seo}, {Shan}, {Shi}, {Siedenburg}, {Son}, {Song}, {Sun}, {Tacconi}, {Tang}, {Tang}, {Tao}, {Tescaro}, {Ting}, {Ting}, {Tomassetti}, {Torsti}, {T{\"u}rko{\v{g}}lu}, {Urban}, {Vagelli}, {Valente}, {Vannini}, {Valtonen}, {V{\'a}zquez Acosta}, {Vecchi}, {Velasco}, {Vialle}, {Vitale}, {Vitillo}, {Wang}, {Wang}, {Wang}, {Wang}, {Wang}, {Wang}, {Wei}, {Weng}, {Whitman}, {Wienkenh{\"o}ver}, {Wu}, {Wu}, {Xia}, {Xiong}, {Xu}, {Yan}, {Yang}, {Yang}, {Yang}, {Yi}, {Yu}, {Yu}, {Zeissler}, {Zhang}, {Zhang}, {Zhang}, {Zhang}, {Zhang}, {Zhang}, {Zheng}, {Zhu}, {Zhuang}, {Zhukov}, {Zichichi}, {Zimmermann}, {Zuccon}, \& {AMS Collaboration}}]{aguilar_bc}
{Aguilar}, M., {Ali Cavasonza}, L., {Ambrosi}, G., {et~al.} 2016, \prl, 117, 231102, \dodoi{10.1103/PhysRevLett.117.231102}

\bibitem[{Aguilar {et~al.}(2021)Aguilar, {Ali Cavasonza}, Ambrosi, Arruda, Attig, Barao, Barrin, Bartoloni, {Başeğmez-du Pree}, Bates, Battiston, Behlmann, Beischer, Berdugo, Bertucci, Bindi, {de Boer}, Bollweg, Borgia, Boschini, Bourquin, Bueno, Burger, Burger, Burmeister, Cai, Capell, Casaus, Castellini, Cervelli, Chang, Chen, Chen, Chen, Cheng, Chou, Chouridou, Choutko, Chung, Clark, Coignet, Consolandi, Contin, Corti, Cui, Dadzie, Dai, Delgado, {Della Torre}, Demirköz, Derome, {Di Falco}, {Di Felice}, Díaz, Dimiccoli, {von Doetinchem}, Dong, Donnini, Duranti, Egorov, Eline, Feng, Fiandrini, Fisher, Formato, Freeman, Galaktionov, Gámez, García-López, Gargiulo, Gast, Gebauer, Gervasi, Giovacchini, Gómez-Coral, Gong, Goy, Grabski, Grandi, Graziani, Guo, Haino, Han, Hashmani, He, Heber, Hsieh, Hu, Huang, Hungerford, Incagli, Jang, Jia, Jinchi, Kanishev, Khiali, Kim, Kirn, Konyushikhin, Kounina, Kounine, Koutsenko, Kuhlman, Kulemzin, {La Vacca}, Laudi, Laurenti, Lazzizzera, Lebedev, Lee, Lee, Leluc,
  Li, Li, Li, Li, Li, Li, Light, Lin, Lippert, Liu, Lu, Lu, Luebelsmeyer, Luo, Lyu, Machate, Mañá, Marín, Marquardt, Martin, Martínez, Masi, Maurin, Menchaca-Rocha, Meng, Mo, Molero, Mott, Mussolin, Ni, Nikonov, Nozzoli, Oliva, Orcinha, Palermo, Palmonari, Paniccia, Pashnin, Pauluzzi, Pensotti, Phan, Plyaskin, Pohl, Porter, Qi, Qin, Qu, Quadrani, Rancoita, Rapin, {Reina Conde}, Rosier-Lees, Rozhkov, Rozza, Sagdeev, Schael, Schmidt, {Schulz von Dratzig}, Schwering, Seo, Shan, Shi, Siedenburg, Solano, Song, Sonnabend, Sun, Sun, Tacconi, Tang, Tang, Tian, Ting, Ting, Tomassetti, Torsti, Tüysüz, Urban, Usoskin, Vagelli, Vainio, Valente, Valtonen, {Vázquez Acosta}, Vecchi, Velasco, Vialle, Wang, Wang, Wang, Wang, Wang, Wang, Wei, Weng, Wu, Xiong, Xu, Yan, Yang, Yi, Yu, Yu, Zannoni, Zhang, Zhang, Zhang, Zhang, Zhang, Zhao, Zheng, Zhuang, Zhukov, Zichichi, Zimmermann, \& Zuccon}]{AGUILAR2021}
Aguilar, M., {Ali Cavasonza}, L., Ambrosi, G., {et~al.} 2021, \phr, 894, 1, \dodoi{https://doi.org/10.1016/j.physrep.2020.09.003}

\bibitem[{{Alemanno} {et~al.}(2024){Alemanno}, {Altomare}, {An}, {Azzarello}, {Barbato}, {Bernardini}, {Bi}, {Cagnoli}, {Cai}, {Casilli}, {Catanzani}, {Chang}, {Chen}, {Chen}, {Chen}, {Coppin}, {Cui}, {Cui}, {Cui}, {Dai}, {de Benedittis}, {de Mitri}, {de Palma}, {Deliyergiyev}, {di Giovanni}, {di Santo}, {Ding}, {Dong}, {Dong}, {Donvito}, {Droz}, {Duan}, {Duan}, {Fan}, {Fan}, {Fang}, {Fang}, {Feng}, {Feng}, {Fernandez Alonso}, {Frieden}, {Fusco}, {Gao}, {Gargano}, {Gong}, {Gong}, {Guo}, {Guo}, {Han}, {Hu}, {Huang}, {Huang}, {Huang}, {Ionica}, {Jiang}, {Jiang}, {Jiang}, {Kong}, {Kotenko}, {Kyratzis}, {Lei}, {Li}, {Li}, {Li}, {Li}, {Liang}, {Liu}, {Liu}, {Liu}, {Liu}, {Liu}, {Loparco}, {Luo}, {Ma}, {Ma}, {Ma}, {Ma}, {Marsella}, {Mazziotta}, {Mo}, {Salinas}, {Niu}, {Pan}, {Parenti}, {Peng}, {Peng}, {Perrina}, {Putti-Garcia}, {Qiao}, {Rao}, {Ruina}, {Shangguan}, {Shen}, {Shen}, {Shen}, {Silveri}, {Song}, {Stolpovskiy}, {Su}, {Su}, {Sun}, {Sun}, {Surdo}, {Teng}, {Tykhonov}, {Wang}, {Wang}, {Wang}, {Wang}, {Wang},
  {Wang}, {Wang}, {Wang}, {Wang}, {Wei}, {Wei}, {Wei}, {Wu}, {Wu}, {Wu}, {Wu}, {Wu}, {Xia}, {Xu}, {Xu}, {Xu}, {Xu}, {Xu}, {Xu}, {Xue}, {Yang}, {Yang}, {Yang}, {Yao}, {Yu}, {Yuan}, {Yuan}, {Yue}, {Zang}, {Zhang}, {Zhang}, {Zhang}, {Zhang}, {Zhang}, {Zhang}, {Zhang}, {Zhang}, {Zhang}, {Zhang}, {Zhao}, {Zhao}, {Zhao}, {Zhou}, {Zhu}, \& {Dampe Collaboration}}]{DAMPE2024}
{Alemanno}, F., {Altomare}, C., {An}, Q., {et~al.} 2024, \prd, 109, L121101, \dodoi{10.1103/PhysRevD.109.L121101}

\bibitem[{{Amato} \& {Blasi}(2018)}]{amato_blasi_18}
{Amato}, E., \& {Blasi}, P. 2018, \adspr, 62, 2731, \dodoi{10.1016/j.asr.2017.04.019}

\bibitem[{{An} {et~al.}(2019){An}, {Asfandiyarov}, {Azzarello}, {Bernardini}, {Bi}, {Cai}, {Chang}, {Chen}, {Chen}, {Chen}, {Chen}, {Cui}, {Cui}, {Dai}, {D'Amone}, {De Benedittis}, {De Mitri}, {Di Santo}, {Ding}, {Dong}, {Dong}, {Dong}, {Donvito}, {Droz}, {Duan}, {Duan}, {D'Urso}, {Fan}, {Fan}, {Fang}, {Feng}, {Feng}, {Fusco}, {Gallo}, {Gan}, {Gao}, {Gargano}, {Gong}, {Gong}, {Guo}, {Guo}, {Guo}, {Han}, {Hu}, {Huang}, {Huang}, {Huang}, {Ionica}, {Jiang}, {Jin}, {Kong}, {Lei}, {Li}, {Li}, {Li}, {Li}, {Li}, {Liang}, {Liang}, {Liao}, {Liu}, {Liu}, {Liu}, {Liu}, {Liu}, {Liu}, {Loparco}, {Luo}, {Ma}, {Ma}, {Ma}, {Ma}, {Ma}, {Marsella}, {Mazziotta}, {Mo}, {Niu}, {Pan}, {Peng}, {Peng}, {Qiao}, {Rao}, {Salinas}, {Shang}, {Shen}, {Shen}, {Shen}, {Song}, {Su}, {Su}, {Sun}, {Surdo}, {Teng}, {Tykhonov}, {Vitillo}, {Wang}, {Wang}, {Wang}, {Wang}, {Wang}, {Wang}, {Wang}, {Wang}, {Wang}, {Wang}, {Wang}, {Wang}, {Wang}, {Wei}, {Wei}, {Wei}, {Wen}, {Wu}, {Wu}, {Wu}, {Wu}, {Wu}, {Xi}, {Xia}, {Xu}, {Xu}, {Xu}, {Xu}, {Xue},
  {Yang}, {Yang}, {Yang}, {Yang}, {Yao}, {Yu}, {Yuan}, {Yue}, {Zang}, {Zhang}, {Zhang}, {Zhang}, {Zhang}, {Zhang}, {Zhang}, {Zhang}, {Zhang}, {Zhang}, {Zhang}, {Zhang}, {Zhang}, {Zhang}, {Zhao}, {Zhao}, {Zhao}, {Zhou}, {Zhou}, {Zhu}, {Zhu}, \& {Zimmer}}]{DAMPE2019}
{An}, Q., {Asfandiyarov}, R., {Azzarello}, P., {et~al.} 2019, \scia, 5, eaax3793, \dodoi{10.1126/sciadv.aax3793}

\bibitem[{Andersen(1954)}]{ANDERSEN1954}
Andersen, E.~S. 1954, Mathematica Scandinavica, 2, 195.
\newblock \url{http://eudml.org/doc/165543}

\bibitem[{{Armillotta} {et~al.}(2025){Armillotta}, {Ostriker}, \& {Linzer}}]{Armillotta2025}
{Armillotta}, L., {Ostriker}, E.~C., \& {Linzer}, N.~B. 2025, arXiv e-prints, arXiv:2507.00120, \dodoi{10.48550/arXiv.2507.00120}

\bibitem[{{Bai}(2022)}]{bai_2021}
{Bai}, X.-N. 2022, \apj, 928, 112, \dodoi{10.3847/1538-4357/ac56e1}

\bibitem[{{Bai} {et~al.}(2019){Bai}, {Ostriker}, {Plotnikov}, \& {Stone}}]{bai_mhd_pic}
{Bai}, X.-N., {Ostriker}, E.~C., {Plotnikov}, I., \& {Stone}, J.~M. 2019, \apj, 876, 60, \dodoi{10.3847/1538-4357/ab1648}

\bibitem[{{Bandyopadhyay} {et~al.}(2020){Bandyopadhyay}, {Yang}, {Matthaeus}, {Chasapis}, {Parashar}, {Russell}, {Strangeway}, {Torbert}, {Giles}, {Gershman}, {Pollock}, {Moore}, \& {Burch}}]{Bandyopadhyay2020}
{Bandyopadhyay}, R., {Yang}, Y., {Matthaeus}, W.~H., {et~al.} 2020, \apjl, 893, L25, \dodoi{10.3847/2041-8213/ab846e}

\bibitem[{{Beattie} {et~al.}(2024){Beattie}, {Federrath}, {Klessen}, {Cielo}, \& {Bhattacharjee}}]{Beattie2024}
{Beattie}, J.~R., {Federrath}, C., {Klessen}, R.~S., {Cielo}, S., \& {Bhattacharjee}, A. 2024, arXiv e-prints, arXiv:2405.16626, \dodoi{10.48550/arXiv.2405.16626}

\bibitem[{{Birdsall} \& {Langdon}(1991)}]{tsc_ref}
{Birdsall}, C., \& {Langdon}, A. 1991, {Plasma Physics via Computer Simulation (1st ed.)}

\bibitem[{{Blasi} \& {Amato}(2012)}]{Blasi_Amato2012}
{Blasi}, P., \& {Amato}, E. 2012, \jcap, 2012, 011, \dodoi{10.1088/1475-7516/2012/01/011}

\bibitem[{{Blasi} {et~al.}(2012){Blasi}, {Amato}, \& {Serpico}}]{blasi12}
{Blasi}, P., {Amato}, E., \& {Serpico}, P.~D. 2012, \prl, 109, 061101, \dodoi{10.1103/PhysRevLett.109.061101}

\bibitem[{Boris(1970)}]{boris}
Boris, J.~P. 1970, Proceeding of Fourth Conference on Numerical Simulations of Plasmas

\bibitem[{{Bouchaud} \& {Georges}(1990)}]{Bouchaud1990}
{Bouchaud}, J.-P., \& {Georges}, A. 1990, \physrep, 195, 127, \dodoi{10.1016/0370-1573(90)90099-N}

\bibitem[{{Butsky} {et~al.}(2024){Butsky}, {Hopkins}, {Kempski}, {Ponnada}, {Quataert}, \& {Squire}}]{Butsky2024}
{Butsky}, I.~S., {Hopkins}, P.~F., {Kempski}, P., {et~al.} 2024, \mnras, 528, 4245, \dodoi{10.1093/mnras/stae276}

\bibitem[{Chandran(2000)}]{chandran_scattering}
Chandran, B. D.~G. 2000, \prl, 85, 4656, \dodoi{10.1103/PhysRevLett.85.4656}

\bibitem[{{Cowsik} \& {Burch}(2010)}]{Burch2010}
{Cowsik}, R., \& {Burch}, B. 2010, \prd, 82, 023009, \dodoi{10.1103/PhysRevD.82.023009}

\bibitem[{{Dampe Collaboration}(2022)}]{DAMPE2022}
{Dampe Collaboration}. 2022, Science Bulletin, 67, 2162, \dodoi{10.1016/j.scib.2022.10.002}

\bibitem[{{Dong} {et~al.}(2022){Dong}, {Wang}, {Huang}, {Comisso}, {Sandstrom}, \& {Bhattacharjee}}]{dong_2022}
{Dong}, C., {Wang}, L., {Huang}, Y.-M., {et~al.} 2022, \scia, 8, eabn7627, \dodoi{10.1126/sciadv.abn7627}

\bibitem[{{Ewart} {et~al.}(2024){Ewart}, {Reichherzer}, {Bott}, {Kunz}, \& {Schekochihin}}]{Ewart2024}
{Ewart}, R.~J., {Reichherzer}, P., {Bott}, A. F.~A., {Kunz}, M.~W., \& {Schekochihin}, A.~A. 2024, \mnras, 532, 2098, \dodoi{10.1093/mnras/stae1578}

\bibitem[{{Farmer} \& {Goldreich}(2004)}]{farmer_goldreich}
{Farmer}, A.~J., \& {Goldreich}, P. 2004, \apj, 604, 671, \dodoi{10.1086/382040}

\bibitem[{{Fielding} {et~al.}(2023){Fielding}, {Ripperda}, \& {Philippov}}]{fielding_plasmoid}
{Fielding}, D.~B., {Ripperda}, B., \& {Philippov}, A.~A. 2023, \apjl, 949, L5, \dodoi{10.3847/2041-8213/accf1f}

\bibitem[{{Fornieri} {et~al.}(2021){Fornieri}, {Gaggero}, {Cerri}, {De La Torre Luque}, \& {Gabici}}]{fornieri_2021}
{Fornieri}, O., {Gaggero}, D., {Cerri}, S.~S., {De La Torre Luque}, P., \& {Gabici}, S. 2021, \mnras, 502, 5821, \dodoi{10.1093/mnras/stab355}

\bibitem[{{Gaensler} {et~al.}(2005){Gaensler}, {Haverkorn}, {Staveley-Smith}, {Dickey}, {McClure-Griffiths}, {Dickel}, \& {Wolleben}}]{Gaensler2005}
{Gaensler}, B.~M., {Haverkorn}, M., {Staveley-Smith}, L., {et~al.} 2005, \sci, 307, 1610, \dodoi{10.1126/science.1108832}

\bibitem[{{Galishnikova} {et~al.}(2022){Galishnikova}, {Kunz}, \& {Schekochihin}}]{alisa_dynamo}
{Galishnikova}, A.~K., {Kunz}, M.~W., \& {Schekochihin}, A.~A. 2022, \prx, 12, 041027, \dodoi{10.1103/PhysRevX.12.041027}

\bibitem[{{Goldreich} \& {Sridhar}(1995)}]{gs95}
{Goldreich}, P., \& {Sridhar}, S. 1995, \apj, 438, 763, \dodoi{10.1086/175121}

\bibitem[{{Grete} {et~al.}(2023){Grete}, {O'Shea}, \& {Beckwith}}]{Grete:2023}
{Grete}, P., {O'Shea}, B.~W., \& {Beckwith}, K. 2023, \apjl, 942, L34, \dodoi{10.3847/2041-8213/acaea7}

\bibitem[{Haverkorn(2015)}]{Haverkorn2015}
Haverkorn, M. 2015, Magnetic Fields in the Milky Way, ed. A.~Lazarian, E.~M. de~Gouveia Dal~Pino, \& C.~Melioli (Berlin, Heidelberg: Springer Berlin Heidelberg), 483--506, \dodoi{10.1007/978-3-662-44625-6_17}

\bibitem[{{Heesen} {et~al.}(2023){Heesen}, {de Gasperin}, {Schulz}, {Basu}, {Beck}, {Br{\"u}ggen}, {Dettmar}, {Stein}, {Gajovi{\'c}}, {Tabatabaei}, \& {Reichherzer}}]{Heesen2023}
{Heesen}, V., {de Gasperin}, F., {Schulz}, S., {et~al.} 2023, \aap, 672, A21, \dodoi{10.1051/0004-6361/202245223}

\bibitem[{{Holcomb} \& {Spitkovsky}(2019)}]{Holcomb2019}
{Holcomb}, C., \& {Spitkovsky}, A. 2019, \apj, 882, 3, \dodoi{10.3847/1538-4357/ab328a}

\bibitem[{{Hopkins} {et~al.}(2022){Hopkins}, {Squire}, {Butsky}, \& {Ji}}]{hopkins_sc_et_problems}
{Hopkins}, P.~F., {Squire}, J., {Butsky}, I.~S., \& {Ji}, S. 2022, \mnras, 517, 5413, \dodoi{10.1093/mnras/stac2909}

\bibitem[{{Hu} {et~al.}(2022){Hu}, {Lazarian}, \& {Xu}}]{hu_2022}
{Hu}, Y., {Lazarian}, A., \& {Xu}, S. 2022, \mnras, 512, 2111, \dodoi{10.1093/mnras/stac319}

\bibitem[{{Huang} \& {Bhattacharjee}(2012)}]{Huang2012}
{Huang}, Y.-M., \& {Bhattacharjee}, A. 2012, \prl, 109, 265002, \dodoi{10.1103/PhysRevLett.109.265002}

\bibitem[{{Jaffe} {et~al.}(2010){Jaffe}, {Leahy}, {Banday}, {Leach}, {Lowe}, \& {Wilkinson}}]{jaffe2010}
{Jaffe}, T.~R., {Leahy}, J.~P., {Banday}, A.~J., {et~al.} 2010, \mnras, 401, 1013, \dodoi{10.1111/j.1365-2966.2009.15745.x}

\bibitem[{{Ji} {et~al.}(2022){Ji}, {Squire}, \& {Hopkins}}]{Ji2022}
{Ji}, S., {Squire}, J., \& {Hopkins}, P.~F. 2022, \mnras, 513, 282, \dodoi{10.1093/mnras/stac895}

\bibitem[{{Kempski} {et~al.}(2023){Kempski}, {Fielding}, {Quataert}, {Galishnikova}, {Kunz}, {Philippov}, \& {Ripperda}}]{Kempski2023}
{Kempski}, P., {Fielding}, D.~B., {Quataert}, E., {et~al.} 2023, \mnras, 525, 4985, \dodoi{10.1093/mnras/stad2609}

\bibitem[{{Kempski} \& {Quataert}(2022)}]{kq2022}
{Kempski}, P., \& {Quataert}, E. 2022, \mnras, 514, 657, \dodoi{10.1093/mnras/stac1240}

\bibitem[{{Kulsrud} \& {Pearce}(1969)}]{kp69}
{Kulsrud}, R., \& {Pearce}, W.~P. 1969, \apj, 156, 445, \dodoi{10.1086/149981}

\bibitem[{{Kulsrud} \& {Cesarsky}(1971)}]{kc71}
{Kulsrud}, R.~M., \& {Cesarsky}, C.~J. 1971, \apjl, 8, 189

\bibitem[{{Lazarian} \& {Xu}(2021)}]{lazarian_xu_mirror}
{Lazarian}, A., \& {Xu}, S. 2021, \apj, 923, 53, \dodoi{10.3847/1538-4357/ac2de9}

\bibitem[{{Lazarian} \& {Yan}(2014)}]{Lazarian2014}
{Lazarian}, A., \& {Yan}, H. 2014, \apj, 784, 38, \dodoi{10.1088/0004-637X/784/1/38}

\bibitem[{{Lemmerz} {et~al.}(2025){Lemmerz}, {Shalaby}, {Pfrommer}, \& {Thomas}}]{Lemmerz2025}
{Lemmerz}, R., {Shalaby}, M., {Pfrommer}, C., \& {Thomas}, T. 2025, \apj, 979, 34, \dodoi{10.3847/1538-4357/ad8eb3}

\bibitem[{{Lemoine}(2023)}]{Lemoine2023}
{Lemoine}, M. 2023, \jopp, 89, 175890501, \dodoi{10.1017/S0022377823000946}

\bibitem[{{Liang} \& {Oh}(2025)}]{Liang2025}
{Liang}, N., \& {Oh}, S.~P. 2025, arXiv e-prints, arXiv:2503.10747, \dodoi{10.48550/arXiv.2503.10747}

\bibitem[{{L{\"u}bke} {et~al.}(2025){L{\"u}bke}, {Reichherzer}, {Aerdker}, {Effenberger}, {Wilbert}, {Fichtner}, \& {Grauer}}]{Lubke2025}
{L{\"u}bke}, J., {Reichherzer}, P., {Aerdker}, S., {et~al.} 2025, arXiv e-prints, arXiv:2505.18155, \dodoi{10.48550/arXiv.2505.18155}

\bibitem[{{Majeski} {et~al.}(2021){Majeski}, {Ji}, {Jara-Almonte}, \& {Yoo}}]{Majeski2021}
{Majeski}, S., {Ji}, H., {Jara-Almonte}, J., \& {Yoo}, J. 2021, \pop, 28, 092106, \dodoi{10.1063/5.0059017}

\bibitem[{{Narayan} \& {Medvedev}(2001)}]{NarayanMedvedev2001}
{Narayan}, R., \& {Medvedev}, M.~V. 2001, \apjl, 562, L129, \dodoi{10.1086/338325}

\bibitem[{{Ohno} \& {Shibata}(1993)}]{Ohno1993}
{Ohno}, H., \& {Shibata}, S. 1993, \mnras, 262, 953, \dodoi{10.1093/mnras/262.4.953}

\bibitem[{{Palyulin} {et~al.}(2019){Palyulin}, {Blackburn}, {Lomholt}, {Watkins}, {Metzler}, {Klages}, \& {Chechkin}}]{Palyulin2019}
{Palyulin}, V.~V., {Blackburn}, G., {Lomholt}, M.~A., {et~al.} 2019, \njp, 21, 103028, \dodoi{10.1088/1367-2630/ab41bb}

\bibitem[{{Petropoulou} {et~al.}(2018){Petropoulou}, {Christie}, {Sironi}, \& {Giannios}}]{Petropoulou2018}
{Petropoulou}, M., {Christie}, I.~M., {Sironi}, L., \& {Giannios}, D. 2018, \mnras, 475, 3797, \dodoi{10.1093/mnras/sty033}

\bibitem[{{Pezzi} \& {Blasi}(2024)}]{Pezzi2024}
{Pezzi}, O., \& {Blasi}, P. 2024, \mnras, 529, L13, \dodoi{10.1093/mnrasl/slad192}

\bibitem[{{Recchia} \& {Gabici}(2024)}]{Recchia2024}
{Recchia}, S., \& {Gabici}, S. 2024, \aap, 692, A20, \dodoi{10.1051/0004-6361/202349005}

\bibitem[{Redner(2001)}]{Redner_2001}
Redner, S. 2001, A Guide to First-Passage Processes (Cambridge University Press)

\bibitem[{{Reichherzer} {et~al.}(2025){Reichherzer}, {Bott}, {Ewart}, {Gregori}, {Kempski}, {Kunz}, \& {Schekochihin}}]{Reichherzer2025}
{Reichherzer}, P., {Bott}, A. F.~A., {Ewart}, R.~J., {et~al.} 2025, \natas, 9, 438, \dodoi{10.1038/s41550-024-02442-1}

\bibitem[{{Ruszkowski} \& {Pfrommer}(2023)}]{Ruszkowski_Pfrommer2023}
{Ruszkowski}, M., \& {Pfrommer}, C. 2023, \aapr, 31, 4, \dodoi{10.1007/s00159-023-00149-2}

\bibitem[{{Schekochihin} {et~al.}(2004){Schekochihin}, {Cowley}, {Taylor}, {Maron}, \& {McWilliams}}]{schekochihin_2004}
{Schekochihin}, A.~A., {Cowley}, S.~C., {Taylor}, S.~F., {Maron}, J.~L., \& {McWilliams}, J.~C. 2004, \apj, 612, 276, \dodoi{10.1086/422547}

\bibitem[{{Schroer} {et~al.}(2025){Schroer}, {Caprioli}, \& {Blasi}}]{Schroer2025}
{Schroer}, B., {Caprioli}, D., \& {Blasi}, P. 2025, \prl, 134, 045201, \dodoi{10.1103/PhysRevLett.134.045201}

\bibitem[{{Shalaby} {et~al.}(2021){Shalaby}, {Thomas}, \& {Pfrommer}}]{shalaby2021}
{Shalaby}, M., {Thomas}, T., \& {Pfrommer}, C. 2021, \apj, 908, 206, \dodoi{10.3847/1538-4357/abd02d}

\bibitem[{{Shalchi}(2019)}]{Shalchi2019}
{Shalchi}, A. 2019, \apjl, 881, L27, \dodoi{10.3847/2041-8213/ab379d}

\bibitem[{{Skilling}(1971)}]{skilling71}
{Skilling}, J. 1971, \apj, 170, 265, \dodoi{10.1086/151210}

\bibitem[{{Stone} \& {Gardiner}(2009)}]{StoneGardiner2009}
{Stone}, J.~M., \& {Gardiner}, T. 2009, \na, 14, 139, \dodoi{10.1016/j.newast.2008.06.003}

\bibitem[{{Stone} {et~al.}(2024){Stone}, {Mullen}, {Fielding}, {Grete}, {Guo}, {Kempski}, {Most}, {White}, \& {Wong}}]{Stone2024}
{Stone}, J.~M., {Mullen}, P.~D., {Fielding}, D., {et~al.} 2024, arXiv e-prints, arXiv:2409.16053, \dodoi{10.48550/arXiv.2409.16053}

\bibitem[{{Strong} {et~al.}(2007){Strong}, {Moskalenko}, \& {Ptuskin}}]{Strong2007}
{Strong}, A.~W., {Moskalenko}, I.~V., \& {Ptuskin}, V.~S. 2007, \arnps, 57, 285, \dodoi{10.1146/annurev.nucl.57.090506.123011}

\bibitem[{{Sun} {et~al.}(2024){Sun}, {Bai}, \& {Zhao}}]{Sun2024}
{Sun}, X., {Bai}, X.-N., \& {Zhao}, X. 2024, arXiv e-prints, arXiv:2409.08592, \dodoi{10.48550/arXiv.2409.08592}

\bibitem[{{Uhlenbeck} \& {Ornstein}(1930)}]{uhlenbeck_ornstein_1930}
{Uhlenbeck}, G.~E., \& {Ornstein}, L.~S. 1930, \pr, 36, 823, \dodoi{10.1103/PhysRev.36.823}

\bibitem[{{Xu} \& {Lazarian}(2018)}]{Xu2018}
{Xu}, S., \& {Lazarian}, A. 2018, \apj, 868, 36, \dodoi{10.3847/1538-4357/aae840}

\bibitem[{{Xu} \& {Lazarian}(2020)}]{xu_2020_trapping}
---. 2020, \apj, 894, 63, \dodoi{10.3847/1538-4357/ab8465}

\bibitem[{{Yan} \& {Lazarian}(2004)}]{yan_lazarian_2004}
{Yan}, H., \& {Lazarian}, A. 2004, \apj, 614, 757, \dodoi{10.1086/423733}

\bibitem[{{Yan} \& {Lazarian}(2008)}]{yan_lazarian_2008}
---. 2008, \apj, 673, 942, \dodoi{10.1086/524771}

\bibitem[{{Yang} {et~al.}(2019){Yang}, {Wan}, {Matthaeus}, {Shi}, {Parashar}, {Lu}, \& {Chen}}]{Yang2019}
{Yang}, Y., {Wan}, M., {Matthaeus}, W.~H., {et~al.} 2019, \pop, 26, 072306, \dodoi{10.1063/1.5099360}

\bibitem[{{Yuen} \& {Lazarian}(2020)}]{Yuen2020}
{Yuen}, K.~H., \& {Lazarian}, A. 2020, \apj, 898, 66, \dodoi{10.3847/1538-4357/ab9360}

\bibitem[{Zweibel(2013)}]{zweibel_micro}
Zweibel, E.~G. 2013, \pop, 20, 055501, \dodoi{10.1063/1.4807033}

\bibitem[{{Zweibel}(2020)}]{zweibel_2020}
{Zweibel}, E.~G. 2020, \apj, 890, 67, \dodoi{10.3847/1538-4357/ab67bf}

\end{thebibliography}
\bibliographystyle{aasjournal}



\end{document}